\documentclass{IEEEtran}
\usepackage{cite,graphicx,subfigure,epsfig,amsmath,amssymb}
\usepackage{amsmath,epsfig,cite}
\usepackage{graphicx,epsfig}
\usepackage{subfigure}
\usepackage{epsfig}
\usepackage{amsmath}
\usepackage{amssymb}

\def \SINR{\textrm{SINR}}

\def \bc{\boldsymbol{c}}
\def \btc{\tilde{\boldsymbol{c}}}
\def \bGamma{\boldsymbol{\Gamma}}
\def \bof{\boldsymbol{f}}

\def \bdelta{\boldsymbol{\delta}}

\def \bDelta{\boldsymbol{\Delta}}

\def \bPhi{\boldsymbol{\Phi}}

\def \bzero{\boldsymbol{0}}
\def \bone{\boldsymbol{1}}

\def \bg{\boldsymbol{g}}
\def \bG{\boldsymbol{G}}
\def \bH{\boldsymbol{H}}

\def \bI{\boldsymbol{I}}

\def \bw{\boldsymbol{w}}

\def \bI{\boldsymbol{I}}
\def \bV{\boldsymbol{V}}
\def \bJ{\boldsymbol{J}}

\def \bLambda{\boldsymbol{\Lambda}}

\def \bh{\boldsymbol{h}}

\def \bSigma{\boldsymbol{\Sigma}}

\def \ba{\boldsymbol{a}}
\def \bb{\boldsymbol{b}}
\def \br{\boldsymbol{r}}

\def \bR{\boldsymbol{R}}

\def \bQ{\boldsymbol{Q}}
\def \bq{\boldsymbol{q}}
\def \bp{\boldsymbol{p}}

\def \E{\textrm{E}}
\def \tr{\textrm{tr}}
\def \diag{\textrm{diag}}

\def \st{\textrm{s.t. }}
\def \Pmax{P_{\textrm{tot}}^{\textrm{max}}}
\def \Ptot{P_{\textrm{tot}}}

\newcommand{\bF}{\boldsymbol{F}}

\newcommand{\bn}{\boldsymbol{n}}
\newcommand{\bB}{\boldsymbol{B}}

\newcommand{\bx}{\boldsymbol{x}}

\newcommand{\bA}{\boldsymbol{A}}
\newcommand{\bW}{\boldsymbol{W}}
\newcommand{\bC}{\boldsymbol{C}}

\newcommand{\bu}{\boldsymbol{u}}

\begin{document}

\markboth{IEEE TRANSACTIONS ON COMMUNICATIONS, VOL. XX, NO. XX, MONTH 2014}{J. Yang \MakeLowercase{\text
it{et al.}}: Joint Design of Optimal Cooperative Jamming and Power Allocation for Linear Precoding}

\title{Joint Design of Optimal Cooperative Jamming and Power Allocation for Linear Precoding}
\author{Jun~Yang,~%\IEEEmembership{Member,~IEEE,}
        Il-Min~Kim,~\IEEEmembership{Senior~Member,~IEEE,}
        Dong~In~Kim,~\IEEEmembership{Senior~Member,~IEEE}
        \thanks{Manuscript received XXX; revised XXX and XXX; accepted XXX. The associate editor coordinating the review of this paper and approving it for publication was XXX.}
        \thanks{This work was supported in part by the Natural Sciences and Engineering Research Council of Canada (NSERC), and in part by National Research Foundation of Korea (NRF) grant funded by the Korean government (MSIP) (2014R1A5A1011478) and the Korea government (MEST) (No. 2012-047720).}
        \thanks{J. Yang is with the Department of Mathematics and Statistics, Queen's University, Kingston, ON K7L 3N6, Canada (e-mail: yangjun@mast.queensu.ca).}
        \thanks{I.-M. Kim is with the Department of Electrical and Computer Engineering, Queen's University, Kingston, ON K7L 3N6, Canada (e-mail: ilmin.kim@queensu.ca).}
        \thanks{D. I. Kim is with the School of Information and Communication Engineering, Sungkyunkwan University (SKKU), Suwon, Korea (e-mail: dikim@skku.ac.kr).}
        }% <-this % stops a space
\maketitle
\begin{abstract}
Linear precoding and cooperative jamming for multiuser broadcast channel is studied to enhance the physical layer security. We consider the system where multiple independent data streams are transmitted from the base station to multiple legitimate users with the help of a friendly jammer. It is assumed that a normalized linear precoding matrix is given at the base station, whereas the power allocated to each user is to be determined. The problem is to jointly design the power allocation across different users for linear precoding and the cooperative jamming at the friendly jammer. The goal is to maximize a lower bound of the secrecy rate, provided that a minimum communication rate to the users is guaranteed. The optimal solution is obtained when the number of antennas at the friendly jammer is no less than the total number of antennas at the users and eavesdropper. Moreover, a suboptimal algorithm is proposed, which can be applied for all the scenarios. Numerical results demonstrate that the proposed schemes are effective for secure communications.
\end{abstract}

\begin{IEEEkeywords}
Cooperative jamming, linear precoding, multiuser broadcast channel, physical layer security.
\end{IEEEkeywords}

\newtheorem{theorem}{Theorem}
\newtheorem{corollary}{Corollary}
\newtheorem{lemma}{Lemma}

\section{Introduction}
\PARstart{E}{nsuring} security of communications at the physical layer has attracted considerable attention in recent years \cite{Li2011,Huang2012,Liao2011,Shi2010,Huang2011a,Dong2010,Goel2008}. Different from the traditional cryptographic algorithms at higher layers, physical layer security exploits the physical characteristics of the wireless transmission medium. For example, secrecy capacity was studied in \cite{Wyner1975,Leung-Yan-Cheong1978,Csiszar1978} from the information-theoretic perspective. Since secrecy capacity is unknown in many cases, the achievable secrecy rate or signal-to-interference-plus-noise ratio (SINR) was also adopted in some work as a metric of security \cite{Li2011,Huang2012,Liao2011,Shi2010,YangSubmitted,YangSubmitted2}.

Physical layer security for multiple antenna systems and/or relay networks has been studied in \cite{Goel2008,Negi2005,Liao2011,Liu2009,Park2013,Liu2013,Goeckel2011}. Among the existing work, the strategy of artificial noise or Cooperative Jamming (CJ) is one of the effective approaches, which was studied by Goel and Negi in \cite{Negi2005,Goel2008} and later by many other researchers \cite{Tekin2008,Liu2008,YangSubmitted,YangSubmitted2}. In most of existing works on CJ, a most typical scenario is that the source transmits only a single data stream to a single legitimate user in the presence of one or multiple eavesdroppers, such as \cite{Liao2011,Li2011,Huang2011a,Huang2012}. In practice, however, multiple independent data streams may be transmitted from the source to multiple legitimate users, such as in multiuser broadcast channels, %\cite{Spencer2004,YangSubmitted,YangSubmitted2},
which has been a very active research topic over the last decade. In the multiuser broadcast channel, the eavesdropper may be interested in any particular stream transmitted by the Base Station (BS). Therefore, it is important to ensure that all the streams from the BS should be kept confidential from the eavesdropper. The zero-forcing approach solely carried out by the BS has major limitations compared to the scheme of using CJ, since it requires the number of antennas at the BS should be no less than the total number of antennas at the eavesdropper and the legitimate users. Also, the power required for zero-forcing approach should be no less than a power budget. Using CJ, the BS can benefit from the friendly jammer since the total instantaneous power could be increases significantly. Also, the CJ can be very effective since the friendly jammer can be selected as the terminals who are close to the eavesdropper but far from the intended receivers.

%However, the secrecy capacity and the optimal CJ for the most general multiuser broadcast channel are still unknown \cite{Fakoorian2011,Liu2010}.
%Recently, the secrecy capacity of multiple input multiple output (MIMO) Gaussian broadcast channel was discussed in \cite{Liu2010} and it was shown that the secrecy capacity region can be achieved by dirty-paper coding (DPC). Due to the high complexity of DPC techniques, however, it is not practical to directly use DPC.

In the literature, the research on practical algorithms for physical layer security in multi-user multi-stream broadcast channels is limited. When the eavesdroppers' channels are known, which is a common assumption in the area of physical layer security \cite{Liao2011,Dong2010,Li2009,Dehghan2012,Fakoorian2011,Jorswieck2008,Zhang2010b}, it was shown in \cite{Liao2011,YangSubmitted,YangSubmitted2} that jointly designing the linear precoding at the BS and the optimal CJ is very difficult\cite{Liao2011,YangSubmitted,YangSubmitted2}. Very recently, in \cite{YangSubmitted,YangSubmitted2}, some optimal CJ algorithms were studied under the assumption that some existing linear precoding/decoding schemes are applied at the BS and the legitimate users. However, the algorithms in \cite{YangSubmitted,YangSubmitted2} are somewhat limited in the sense that the linear precoding matrix at the base station is totally independent of the CJ, meaning that no joint optimization between the BS and the friendly jammer is considered at all. However, fully joint design of linear precoding matrix and the CJ is very difficult. Actually, even in the case of conventional non-secure communications with \emph{no} security conditions or \emph{no} eavesdropper, deriving truly optimal linear precoding matrix is generally very difficult and remains as an open problem. Addressing such shortcoming, in this paper, we investigate joint designing of the CJ and the power allocation of linear precoding matrix.

In this paper, we assume that the BS is able to collect the channel information associated with the users, with which the BS can pre-determine a normalized linear precoding matrix except an individual power allocation to each user. Then the power allocation is jointly optimized with the CJ. Also, we assume the eavesdropper who has multiple antennas could maximize the SINR for each data stream using optimal receive beamforming\cite{Liao2011,Pinto2009,Zhou2011,Pinto2012,Koyluoglu2010a,Wang2011a}.
%As in many existing work on linear precoding matrix design \cite{Peel2005,Schubert2004,Schubert2005,Bengtsson1999,Huang2010,Huang2010b}, we also use the SINR as a metric of quality-of-service (QoS) for each user, and
We assume that each user has one antenna, and the eavesdropper is the legitimate terminal who is currently unscheduled in the downlink. Thus, the channel of eavesdropper is assumed known to the friendly jammer since the eavesdropper is actually an active node in the wireless network whose channel can be monitored. In the area of physical layer security, this is a widely adopted common assumption \cite{Liao2011,Dong2010,Li2009,Dehghan2012,Fakoorian2011,Jorswieck2008,Zhang2010b}.

\emph{Notation:} $(\cdot)^H$ denotes the operator of conjugate transpose and $\E[\cdot]$ is the expectation operator. For positive Hermitian matrix, $(\cdot)^{\frac{1}{2}}$ denotes the Hermitian squared root.
$\bzero_{N\times M}$ denotes an $N\times M$ matrix with all zero elements; $\bI_N$ denotes an $N\times N$ diagonal matrix with diagonal elements equal to one; and
$\mathbb{C}$ denotes the set of complex numbers. Moreover, we use $A:=B$ to denote that $A$ by definition equals to $B$, and use $A=:B$ to denote that $B$ by definition equals to $A$.
%The notation $\veco(\cdot)$ denotes the vectorization of a matrix, i.e., for $\bA=[\ba_1,\cdots,\ba_N]$ where $\ba_i, i=1,\cdots,N$ are column vectors of $\bA$, $\veco(\bA)$ is a vector that equals to $[\ba_1^H,\cdots,\ba_N^H]^H$.
The notation $\|\cdot\|$ denotes the Frobenius norm, and $\|\cdot\|_1$ denotes the $L_1$ norm. Furthermore, the curled inequality symbols $\preceq$ and $\succeq$ (and their strict forms $\prec$ and $\succ$) are used to denote generalized inequalities: between vectors,
they represent componentwise inequalities; between Hermitian matrices, they represent
matrix inequalities.
Finally, for two matrices $\bA\in\mathbb{C}^{N\times N}$ and $\bB\in\mathbb{C}^{M\times M}$, $\diag\{\bA,\bB\}$ denotes the matrix $\left[\begin{array}{cc}\bA & \bzero_{N\times M}\\ \bzero_{M\times N} & \bB\end{array}\right]$.

\section{System Model and Problem Formulation}\label{section_sys_model_pro_formu}
\subsection{System Model}
\begin{figure}
\centering\includegraphics[width=0.37\textwidth]{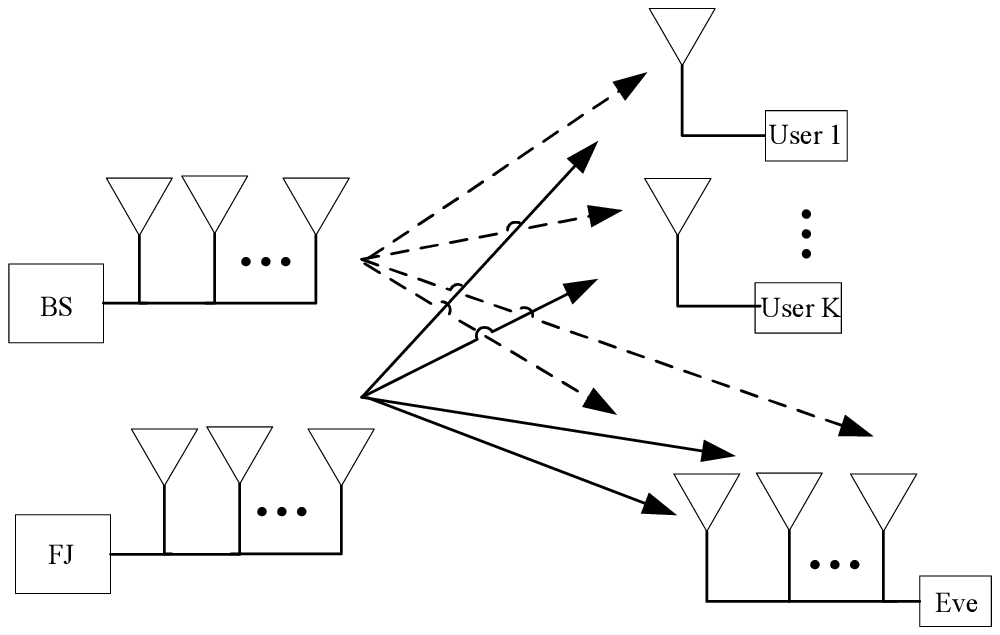}
\caption{System model.}\label{fig0}
\end{figure}
We consider a multiuser broadcast channel as shown in Fig. \ref{fig0}, in which the BS transmits $K$ independent\footnote{This is widely adopted assumption for multi-user broadcast channels.} data streams to $K$ users, each of whom has a single antenna. We assume the BS, the friendly jammer (FJ), and the eavesdropper (Eve) have $N$, $L$, and $Z$ antennas, respectively. The channels from the BS to the users, the BS to Eve, the FJ to the users, and the FJ to Eve are denoted by $\bF=[\bof_1,\cdots,\bof_K]\in\mathbb{C}^{N\times K}$, $\bH=[\bh_1,\cdots,\bh_Z]\in\mathbb{C}^{N\times Z}$, $\bB=[\bb_1,\cdots,\bb_K]\in\mathbb{C}^{L\times K}$, and $\bG=[\bg_1,\cdots,\bg_Z]\in\mathbb{C}^{L\times Z}$, respectively.
The CJ is composed of several independent noises and it is denoted by $\bJ(t)=\sum_{j=1}^{Z}\bq_j z_j(t)$, where $\bq_j$ denotes the weight vector for the $j$-th noise and $z_j(t)$ is the $j$-th independently generated Gaussian noise with zero mean and $\E[|z_j(t)|^2]=1$. Let $\bW$ denote the precoding matrix used at the BS, which is designed for transmitting multiple data streams to multiple users with single receive antennas. In the case of conventional communications with {\it no} security conditions or {\it no} eavesdropper, there are many different ways to design $\bW$. For example, $\bW$ can be obtained in closed-form based on zero-forcing or minimum mean squared error (MMSE) criterions \cite{Peel2005}. Or, $\bW$ might be optimized while guaranteeing the QoS requirements of the users. Unfortunately, in most scenarios where the users' QoS constraints are given, deriving truly optimal $\bW$ is generally very difficult and optimal solutions are generally unknown\cite{Bengtsson1999,Rashid-Farrokhi1998,Wiesel2006}.\footnote{Typically, only some iterative optimization methods were proposed, which are not necessarily provide the truly optimal performance\cite{Wiesel2006}.}

For secure communications, there might be few different approaches in determining $\bW$. A simplest approach is to design $\bW$ simply as in the conventional non-secure communications. A clear benefit is that one can utilize the existing results in the literature. In this approach, however, the security issue or jamming the eavesdropper is totally up to the CJ only (i.e., $\bJ(t)$), with no coordination with precoder $\bW$. Thus, the overall performance can be limited. This approach was used in \cite{YangSubmitted,YangSubmitted2}.
The other extreme approach is that one tries to perfectly carry out joint optimization of $\bW$ and $\bJ(t)$.  If such optimization were doable, a clear benefit would be as follows: the system could be perfectly optimized and the security issue would be addressed by joint optimal coordination of $\bW$ and $\bJ(t)$.
Unfortunately, this approach is analytically intractable in general. In fact, as discussed above, even optimizing only $\bW$ for the conventional (non-secure) communications is generally very difficult when the users' QoS constraints are given.

In this paper, we attempt a balanced approach between the two extremes. Specifically, we carry out {\it partial} joint optimization of $\bJ(t)$ and $\bW$. To this end, we first rewrite $\bW$ as $\bW=[\sqrt{p_1}\bu_1,\cdots,\sqrt{p_K}\bu_K]$, where $\{\|\bu_k\|=1: k=1,\cdots,K\}$. It is easy to see that $p_k$ can be interpreted as the power allocated to the $k$-th user, and $\bu_k$ can be interpreted as the normalized precoding vector designed for the $k$-th user.\footnote{The expression $\bW=[\sqrt{p_1}\bu_1,\cdots,\sqrt{p_K}\bu_K]$ has been used in many existing works in the non-secure communication to design $\bW$, such as in \cite{Yang1998,Rashid-Farrokhi1998,Schubert2004,Schubert2005}. For example, in \cite{Yang1998}, the power assignment problem was considered to design $\{p_k: k=1,\cdots,K\}$ given $\{\bu_k: k=1,\cdots,K\}$. In \cite{Rashid-Farrokhi1998,Schubert2004,Schubert2005}, alternating optimizing $\{\bu_k: k=1,\cdots,K\}$ and $\{p_k: k=1,\cdots,K\}$ were studied.} In this paper, we will carry out joint optimization of the power allocation $\{p_k \}$ and CJ $\bJ(t)$. For $\{\bu_k\}$, one can use any existing results derived for the non-secure communications. Compared to the naive approach (with no joint optimization as in \cite{YangSubmitted,YangSubmitted2}), in our approach, the security issue is addressed by joint optimal coordination of $\{ p_k\}$ and $\bJ(t)$. Thus,  our approach outperforms the naive approach, which will be numerically demonstrated in Section IV. Compared to the full joint optimization of $\bJ(t)$ and $\bW$, which seems analytically intractable, our approach is analytically tractable.

Note that if $L<Z$, the degrees of freedom (DoF) at Eve is larger than the DoF at FJ. Then it is always possible for Eve to cancel any jamming signal transmitted by FJ. In order to ensure that CJ be an effective approach, we will always assume $L\ge Z$ throughout this paper and this assumption will not be explicitly stated in what follows.

\subsection{Problem Formulation}
%In this paper, we will use SINR as a performance metric for the legitimate users and a security metric for Eve\footnote{This is widely adopted and used in the literature\cite{Li2011,Huang2012,Liao2011,Shi2010,YangSubmitted,YangSubmitted2}.}.
The SINR of the $k$-th stream at the $k$-th user can be written as
\begin{equation}\label{}
\begin{split}
    \SINR_k(\bp,\bJ(t))%&:=\frac{|\bof_k^H\bw_k|^2}{\sum_{i\neq k}|\bof_k^H\bw_i|^2+\E[|\bb_k^H\bJ(t)|^2]+\sigma^2}\\
    &=\frac{p_k|\bof_k^H\bu_k|^2}{\sum_{i\neq k}p_i|\bof_k^H\bu_i|^2+\bb_k^H\bSigma\bb_k+\sigma^2}\\
    &:=\SINR_k(\bp,\bSigma),
\end{split}
\end{equation}
where $\bSigma:=\sum_{j=1}^{Z}\bq_j\bq_j^H\in\mathbb{C}^{L\times L}$ is the covariance matrix of CJ\footnote{The number of $\bq_j$ is $Z$ because the expression of $\SINR^U_{e,k}$ is a function of $\bSigma$ only through $\bG^H\bSigma\bG$, which is a $Z\times Z$ matrix.}, and $\sigma^2$ is the noise variance at the users. Note that the SINR depends on $\bJ(t)$ only through $\bSigma$. This means that the design of $\bJ(t)$ can be reduced to the design of $\bSigma$. Thus, we will use notation
$\SINR_k(\bp,\bSigma)$ rather than $\SINR_k(\bp,\bJ(t))$.
In order to guarantee reliable transmission to each user, we design the power allocation vector $\bp=[p_1,p_2,\cdots,p_K]^T$ and the CJ, $\bJ(t)$, such that the communication rate to user $k$ is larger than a given rate threshold, i.e.,
$C_k:=\log\left(1+\SINR_k(\bp,\bSigma)\right)\ge C=:\log(1+\tau)$,
where $C$ is the rate threshold and $\tau$ is the corresponding QoS threshold for each user. On Eve's side, using her multiple antennas, it is possible for Eve to maximize the output SINR of the $k$-th stream using optimal receive beamforming, $\tilde{\bw}_k=\left(\bH^H\bW\bW^H\bH+\sigma^2\bI+\bG^H\bSigma\bG\right)^{-1}\bH^H\bu_k$. The output SINR can be written as
\begin{equation}\label{normal_SINR}
\begin{split}
    &\SINR_{e,k}(\bp,\bSigma)
   :=\\
   &\frac{|\sqrt{p_k}\tilde{\bw}_k^H\bH^H\bu_k|^2}{\tilde{\bw}_k^H\left(\sum_{i\neq k}^K p_i\bH^H\bu_i\bu_i^H\bH+\sigma^2\bI+\bG^H\bSigma\bG\right)\tilde{\bw}_k}=\\
    &\frac{p_k\bu_k^H\bH^H\left(\sum_{i=1}^K     p_i\bH^H\bu_i\bu_i^H\bH+\sigma^2\bI+\bG^H\bSigma\bG\right)^{-1}\bH\bu_k}{1-p_k\bu_k^H\bH^H\left(\sum_{i=1}^K p_i\bH^H\bu_i\bu_i^H\bH+\sigma^2\bI+\bG^H\bSigma\bG\right)^{-1}\bH\bu_k}.
\end{split}
\end{equation}
Note that in the above expression of SINR, the other $(K-1)$ streams except the particular $k$-th stream are considered as interferences when Eve tries to decode the $k$-th stream.

A possible optimization problem is to maximize the minimum secrecy rate under a total power constraint\footnote{Note that individual power constraint of the BS and the jammer might also be of interest, which will be considered in further work.} of linear precoding and CJ, and constraints on the minimum rates to the users:
\begin{equation}\label{very_original_problem}
\begin{split} \max_{\bp,\bSigma}\{\min_{k}C_{se,k}\}
  \quad\st &\sum_{k=1}^K p_k+\tr(\bSigma)\le\Ptot,\quad
   C_k\ge C,\\
   &\quad p_k\ge 0,\quad k=1,\cdots,K,
\end{split}
\end{equation}
where $\small C_{se,k}=\left[\log(1+\SINR_k(\bp,\bSigma))-\log(1+\SINR_{e,k}(\bp,\bSigma))\right]^+$ is the secrecy rate for the $k$-th user's data stream and $\Ptot$ denotes the maximum available power for both FJ and BS. This problem (\ref{very_original_problem}) is generally very difficult to solve because it is non-convex. For analytical tractability, we obtain a lower-bound of the secrecy rate and use it as the cost function. To this end, we first consider an upper bound of $\SINR_{e,k}(\bp,\bSigma)$ as
\begin{equation}\label{worst_case_SINR}
\begin{split}
    &\SINR^U_{e,k}(p_k,\bSigma)=\\
    &\frac{p_k\bu_k^H\bH^H\left(     p_k\bH^H\bu_k\bu_k^H\bH+\sigma^2\bI+\bG^H\bSigma\bG\right)^{-1}\bH\bu_k}{1-p_k\bu_k^H\bH^H\left( p_k\bH^H\bu_k\bu_k^H\bH+\sigma^2\bI+\bG^H\bSigma\bG\right)^{-1}\bH\bu_k},
\end{split}
\end{equation}
where it is easy to prove that $\SINR_{e,k}(\bp,\bSigma)\le \SINR^U_{e,k}(p_k,\bSigma)$
and the equality holds when $\sum_{i=1}^K p_i\bH^H\bu_i\bu_i^H\bH=p_k\bH^H\bu_k\bu_k^H\bH$.
Using the upper bound $\SINR_{e,k}^U(\bp,\bSigma)$, it is possible to obtain
a lower bound of the achievable secrecy rate: $ C_{se,k}\ge C_{se,k}^{\textrm{L,1}}$,
%\begin{equation}\label{temp_lower_bound_secrecy_rate}
%  C_{se,k}\ge C_{se,k}^{\textrm{L,1}},
%\end{equation}
where \begin{equation} C_{se,k}^{\textrm{L,1}}=\left[\log(1+\SINR_k(\bp,\bSigma))
-\log(1+\SINR^U_{e,k}(p_k,\bSigma))\right]^+. \end{equation} If $C_{se,k}^{\textrm{L,1}}$ is used as the cost function, the optimization problem is given by
\begin{equation}\label{equivalent_problem}
\begin{split} \max_{\bp,\bSigma}\{\min_{k}C_{se,k}^{\textrm{L,1}}\}
  \quad\st &\sum_{k=1}^K p_k+\tr(\bSigma)\le\Ptot,\quad
   C_k\ge C,\\
   &\quad p_k\ge 0,\quad k=1,\cdots,K.
\end{split}
\end{equation} Unfortunately, this problem is still difficult to solve in general. Thus, we lower-bound $C_{se,k}^{\textrm{L,1}}$ again. Specifically, from $C_k=\log(1+\SINR_k(\bp,\bSigma))\ge C$, we have $C_{se,k}^{\textrm{L,1}} \geq C_{se,k}^{\textrm{L,2}}$, where
$C_{se,k}^{\textrm{L,2}}=[C-\log(1+\SINR_{e,k}^U(p_k,\bSigma))]^+$. When $C_{se,k}^{\textrm{L,2}}$ is used as the cost function, the optimization problem is given by
\begin{equation}\label{solvable_problem}
\begin{split}
\max_{\bp,\bSigma}\{\min_{k}C_{se,k}^{\textrm{L,2}}\}
    \quad
    \st&\quad\|\bp\|_1+\tr(\bSigma)\le\Ptot,
    \quad C_k\ge C,\\
    &\quad p_k\ge 0,\quad k=1,\cdots,K.
\end{split}
\end{equation}
Finally, from $\max_{\bp,\bSigma}\{\min_{k}C_{se,k}^{\textrm{L,2}}\}=\left[C-\min_{\bp,\bSigma}\max_{k}\log(1+\SINR^U_{e,k}(p_k,\bSigma))\right]^+$, the problem (\ref{solvable_problem}) is equivalent to the following:
\begin{equation}\label{solvable_problem_SINR}
\begin{split}
  \min_{\bp\succeq\bzero, \bSigma}& \left\{\max_k \SINR^U_{e,k}(p_k,\bSigma)\right\}
   \quad\\
    \st&\quad\|\bp\|_1+\tr(\bSigma)\le\Ptot,\\
    &\quad \SINR_k(\bp,\bSigma)\ge\tau,\quad k=1,\cdots,K.
\end{split}
\end{equation}
%Note that since $C_{se,k}^{\textrm{L,1}}\ge C_{se,k}^{\textrm{L,2}}$, the objective function in (\ref{equivalent_problem}) is more tight than (\ref{solvable_problem}).
In the rest of the paper, we focus on solving the problem (\ref{solvable_problem}) or its equivalent form (\ref{solvable_problem_SINR}). We will later show that when $L\ge K+Z$, the solution to (\ref{solvable_problem}) is also the solution to (\ref{equivalent_problem}). Unfortunately, the optimization problems (\ref{solvable_problem}) and (\ref{solvable_problem_SINR}) are still non-convex since both $\SINR^U_{e,k}(p_k,\bSigma)$ and $\SINR_k(\bp,\bSigma)$ are non-convex functions. Thus, it is generally not possible to directly solve (\ref{solvable_problem}) or (\ref{solvable_problem_SINR}). In the next section, the solutions to  (\ref{solvable_problem}) or (\ref{solvable_problem_SINR}) are studied.

\emph{Remark:} In the sense of detection error probability, the optimal strategy for Eve is the maximum likelihood (ML) detection. However, due to the nonlinearity of ML detection, directly analyzing ML detection is very difficult. In this paper, instead of the ML detection, we assume Eve uses beamforming, which is optimal in the sense of maximizing the SINR. Then a lower bound of the secrecy rate based on the SINR upper bound is maximized, which is equivalent to minimizing the SINR upper bound. An interesting question is, ``Which gives better performance for Eve?" Let $P_s^{\textrm{U-SINR}}$ denote the symbol error rate (SER) when the optimal receive beamforming to maximize the upper bound of the SINR is used, and $P_s^{\textrm{ML}}$ denote the SER for ML detection. We can show that $P_s^{\textrm{ML}}\ge P_s^{\textrm{U-SINR}}$. That is, using the upper bound of the SINR is even more conservative than ML dedetection. The proof is given in Appendix \ref{proof_remark}.

\section{Optimal Power Allocation and Cooperative Jamming}\label{section_main_results}
In this section, we investigate the solution to problem (\ref{solvable_problem}). Specifically, we first give the necessary and sufficient condition for the existence of the solution to problem (\ref{solvable_problem}). Then we derive the optimal solution to (\ref{solvable_problem}) when $L\ge K+Z$. Finally, we propose an alternating algorithm based on an asymptotic approximation to get a suboptimal solution to (\ref{solvable_problem}), which does not require the condition $L\ge K+Z$.
%Thus, the proposed alternating algorithm can be applied whether $L$ is larger than $K+Z$ or not.
\subsection{Condition for Existence of Solution}
The solution to (\ref{solvable_problem}) may not exist since the constraints $\{\SINR_k(\bp,\bSigma)\ge\tau: k=1,\cdots,K\}$ may not be satisfied with any $\bp$ and $\bSigma$. Thus, studying the condition that the solution exists is particularly important. In the following lemma, the necessary and sufficient condition for the existence of the solution is given.
\begin{lemma}\label{lemma1}
The solution to (\ref{solvable_problem}) exists if and only if
\begin{equation}\label{existence_condition}
-\sigma^2\left(\bDelta^H\right)^{-1}\bone_{K\times 1}\succeq \bzero\quad \textrm{and}\quad \|-\sigma^2\left(\bDelta^H\right)^{-1}\bone_{K\times 1}\|_1\le\Ptot,
\end{equation}
where the $k$-th column of $\bDelta\in\mathbb{C}^{K\times K}$ is defined as
\begin{equation}\label{}\small
\left[|\bof_k^H\bu_1|^2,\cdots,|\bof_k^H\bu_{k-1}|^2, -\frac{|\bof_k^H\bu_k|^2}{\tau},|\bof_k^H\bu_{k+1}|^2,\cdots,|\bof_k^H\bu_K|^2\right]^H.
\end{equation}
\end{lemma}
\begin{IEEEproof}
See Appendix \ref{proof_lemma1}.
\end{IEEEproof}
The condition given by (\ref{existence_condition}) can be intuitively explained as follows: For given $\bp$, since $\SINR_k(\bp,\bSigma)$ is maximized when $\bSigma=\bzero$, the solution of (\ref{solvable_problem}) exists if and only if there exists $\bp$ satisfying $\|\bp\|_1\le\Ptot$ and $\SINR_k(\bp)\ge\tau$ for all $k$, which are actually the constraints in (\ref{solvable_problem}) when no CJ is transmitted. The existence condition given by (\ref{existence_condition}) is equivalent to the existence for $\bp$ that satisfies both $\|\bp\|_1\le\Ptot$ and $\SINR_k(\bp)\ge\tau$ for all $k$.

From Lemma \ref{lemma1}, one can know that the optimal solution exists if and only if (\ref{existence_condition}) is satisfied.
However, with the condition (\ref{existence_condition}), the problem (\ref{solvable_problem}) is still non-convex and solving the non-convex problem is still very difficult. In the following subsection, we first derive a necessary condition for $\bSigma$ to be optimal when $L\ge K+Z$ and this condition turns out to be very useful to obtain the actual optimal solution when $L\ge K+Z$.

\subsection{Optimal Solution for $L\ge K+Z$}
In this subsection, we solve the problem (\ref{solvable_problem}) when $L\ge K+Z$. We first derive a very important condition for the optimality of CJ's covariance matrix $\bSigma$. Specifically, it turns out that designing CJ to be orthogonal to the users' channel is optimal when $L\ge K+Z$. The result is given in the following lemma.
\begin{lemma}\label{lemma2}
When $L\ge K+Z$ and the condition of (\ref{existence_condition}) is satisfied, the solution $\bSigma_{\textrm{opt}}$ to problem (\ref{solvable_problem}) must be orthogonal to the users' channels, which means $\bB^H\bSigma_{\textrm{opt}}=\bzero_{K\times L}$.
\end{lemma}
\begin{IEEEproof}
See Appendix \ref{proof_lemma2}.
\end{IEEEproof}
Note that  in the existing literature for CJ design, designing CJ such that it has nulls at the users, i.e., zero-forcing condition, is generally suboptimal (rather than optimal)\cite{Dong2010,Li2011}. The result of Lemma \ref{lemma2} shows the if the jammer has enough DoF, the best scheme for the CJ to do is to jam the eavesdropper without interfering the users since the jammer cannot help the legitimate users.

In the following theorem, we show that using the result of Lemma \ref{lemma2}, it is possible to transform the non-convex problem (\ref{solvable_problem}) to a convex problem, which can be readily solved.

\begin{theorem}\label{theorem1}
When $L\ge K+Z$ and the condition of (\ref{existence_condition}) is satisfied, the optimal power allocation vector, $\bp_{\textrm{opt}}$, is given by
\begin{equation}\label{optimal_p}
    \bp_{\textrm{opt}}=-\sigma^2\left(\bDelta^H\right)^{-1}\bone_{K\times 1}\succeq \bzero,
\end{equation}
and the optimal CJ is obtained by $\bSigma_{\textrm{opt}}=\bGamma^H_{\textrm{opt}}\bGamma_{\textrm{opt}}$, where
\begin{equation}\label{optimal_Gamma}
    \bGamma^H_{\textrm{opt}}=\left[
                \begin{array}{cc}
                  \bG & \bB \\
                \end{array}
              \right]\left[
                       \begin{array}{cc}
                         \bG^H\bG & \bG^H\bB \\
                         \bB^H\bG & \bB^H\bB \\
                       \end{array}
                     \right]^{-1}\left[
                                                             \begin{array}{c}
                                                               \bLambda^{1/2} \\
                                                               \bzero_{K\times Z} \\
                                                             \end{array}
                                                           \right]\in\mathbb{C}^{L\times Z},
\end{equation}
in which
\begin{equation}\label{optimal_Lambda}
    \bLambda^{1/2}=\diag\{\sqrt{{x_1}^{-1}-\sigma^2},\cdots,\sqrt{{x_Z}^{-1}-\sigma^2}\}\in\mathbb{C}^{Z\times Z}.
\end{equation}
Denoting new variable $\eta:=\max_k\{\SINR^U_{e,k}(p_k,\bSigma)\}$, the vector $\bx=[x_1,\cdots,x_Z]^T$ is the solution to the following convex optimization problem:
\begin{equation}\label{optimal_lambda}
    \begin{split}
    \bx=&\arg\min_{\bzero_{Z\times 1}\prec \bx\preceq\frac{1}{\sigma^2}\bone_{Z\times 1},\eta}\quad\eta\quad\\
    \st&\quad\sum_{j=1}^{Z}\phi_j{x_j}^{-1}\le\Ptot+\sigma^2\sum_{j=1}^{Z}\phi_j-\|\bp_{\textrm{opt}}\|_1\\
    &\quad p_k\sum_{j=1}^{Z}|a_{kj}|^2 x_j\le\eta, \quad k=1,\cdots,K.
%    &\quad\bzero_{Z\times 1}\prec \bx\preceq\frac{1}{\sigma^2}\bone_{Z\times 1}.
\end{split}
\end{equation}
where $\ba_{k}=[a_{k1},a_{k2},\cdots,a_{kZ}]^T:=\bH^H\bu_k\in\mathbb{C}^{Z\times 1}$ and $\phi_j$ is defined as the $j$-th diagonal element of
$\left[\bG^H\bG-\bG^H\bB\left(\bB^H\bB\right)^{-1}\bB^H\bG\right]^{-1}$.
\end{theorem}
\begin{IEEEproof}
See Appendix \ref{proof_theorem1}.
\end{IEEEproof}
In Theorem \ref{theorem1}, the optimal power allocation, $\bp_{\textrm{opt}}$, for linear precoding can be computed in closed form by (\ref{optimal_p}), and the optimal CJ $\bSigma_{\textrm{opt}}$ can be computed in partially closed form by (\ref{optimal_Gamma}) and (\ref{optimal_Lambda}), where $x_j$ are readily obtained by solving the convex optimization problem of (\ref{optimal_lambda}) numerically, e.g., using the interior-point method. The proposed optimal algorithm can also be implemented distributively, i.e., $\bp_{\textrm{opt}}$ can be computed by the BS using only the information of $\bF$ and then be transmitted to the CJ. The CJ does not need to know $\bF$. After receiving $\bp_{\textrm{opt}}$, the optimal CJ can be designed.

Finally, in the following lemma, we prove that the two problems in (\ref{solvable_problem}) and (\ref{equivalent_problem}) are equivalent when $L\ge K+Z$, i.e., DoF at the FJ is equal to or larger than the total DoF at the legitimate users and Eve.

\begin{lemma}\label{lemma_equivalent}
If $L\ge K+Z$, the problems of (\ref{solvable_problem}) and (\ref{equivalent_problem}) are equivalent.
\end{lemma}
\begin{IEEEproof}
See Appendix \ref{appen_equivalence}.
\end{IEEEproof}

\subsection{Suboptimal Solution}
Note that the optimal algorithm given by Theorem \ref{theorem1} requires the condition that $L\ge K+Z$. If $L<K+Z$, the inversion in (\ref{optimal_Gamma}) does not exists since the matrix $\left[\bG,\bB\right]\in\mathbb{C}^{L\times(Z+K)}$ does not have full row rank. Thus, the main limitation of the optimal algorithm in Theorem \ref{theorem1} is that it cannot be applied when $L<K+Z$. Also, note that the condition $\bB^H\bSigma=\bzero$ in Lemma \ref{lemma2} is no longer a necessary condition for optimality of $\bSigma$ in the case of $L<K+Z$, which can be intuitively explained as follows. To make the condition $\bB^H\bSigma=\bzero$ satisfied, $K$ DoF have been used for the FJ. Then the residual DoF at the FJ to design CJ are just $(L-K)$, which are less than $Z$ when $L<K+Z$. In this case, Eve can easily null any CJ since Eve has more DoF. Thus, the CJ is not effective anymore by $\bB^H\bSigma=\bzero$ when $L<K+Z$. This result is consistent with what is known in the literature, i.e., zero-forcing is not optimal in general. Consequently, in the case of $L<K+Z$, the CJ should be designed such that some power of jamming signal is leaked to the users in order to effectively interfere Eve, rather than zero-forcing. Unfortunately, the optimal solution to (\ref{solvable_problem}) when $L<K+Z$ is very difficult to obtain, because it is non-convex.

In this subsection, we propose a suboptimal algorithm that does not require the condition $L\ge K+Z$, which means the suboptimal algorithm can be always used whether $L$ is greater than $K+Z$ or not. The proposed suboptimal solution is based on alternating algorithms. %\cite{Csisz1984,Yeung2008,Grippo2000,Grippo1999}.
Note that the well-known expectation-maximization (EM) algorithm %\cite{Dempster1977}
and iterative water-filling algorithm \cite{Yu2004}
are examples of the alternating optimization algorithms. In particular, the alternating optimization method is a common approach to handle non-convex problems \cite{Fang2006,Mo2009,Tseng2010,Schubert2004}.%,Tan2011,Niesen2009,Razaviyayn2012}.

The first step is to reformulate the problem (\ref{solvable_problem}) as an equivalent optimization problem.
We therefore consider its equivalent problem (\ref{solvable_problem_SINR}). Since the rank of $\bSigma$ is $Z$, we can always write\footnote{Let the eigenvalue decomposition of $\bSigma$ be $\bSigma=\tilde{\bV}\diag\{\tilde{\bLambda},\bzero_{(L-Z)\times (L-Z)}\}\tilde{\bV}^H$, where $\tilde{\bLambda}$ is a $Z\times Z$ diagonal matrix. Then we can get $\bGamma=[\tilde{\bLambda}^{\frac{1}{2}},\bzero_{Z\times (L-Z)}]\bV^H\in\mathbb{C}^{Z\times L}$.} $\bSigma=\bGamma^H\bGamma$ where $\bGamma\in\mathbb{C}^{Z\times L}$.
Moreover, if we define $\bc_k:=\bGamma\bb_k$, then $\|\bc_k\|^2=\bb_k^H\bSigma\bb_k$ is the amount of CJ power received by the $k$-th user. Using this notation, the optimal $\bp$ and $\bSigma$ of problem (\ref{solvable_problem_SINR}) can be denoted as functions of $\bGamma$ and $\bc_k$ as $\bp(\{\bc_k\})=-(\bDelta^H)^{-1}[\|\bc_1\|+\sigma^2,\cdots,\|\bc_K\|+\sigma^2]^T$ and $\bSigma=\bGamma^H\bGamma$. The optimal $\{\bc_k: k=1,\cdots,K\}$ and $\bGamma$ are obtained by the following non-convex optimization problem:
\begin{equation}\label{alternating_problem}
  \begin{split}
    &\min_{\bGamma,\{\bc_k\},\{x_j\},\eta}\eta\quad\\
      &\st\quad\bG^H\bGamma^H=[\bLambda^{1/2},\bzero]^T\bV^H,\quad \bb_k^H\bGamma^H=\bc_k^H,\quad k=1,\cdots,K,\\
      &\bLambda^{1/2}=\diag\{x_1,x_2,\cdots,x_{Z}\},\quad x_j\ge 0,\quad j=1,\cdots,Z,\\
      &\tr\{\bGamma^H\bGamma\}-\|(\bDelta^H)^{-1}[\|\bc_1\|+\sigma^2,\cdots,\|\bc_K\|+\sigma^2]^T\|_1\le\Ptot,  \\
      &\bdelta_k^H[\|\bc_1\|+\sigma^2,\cdots,\|\bc_K\|+\sigma^2]^T\sum_{j=1}^{Z}\frac{|a_{kj}|^2}{\sigma^2+x_j^2}\le\eta, \quad k=1,\cdots,K,
      %&\bDelta^H\bp+[\|\bc_1\|^2,\cdots,\|\bc_K\|^2]^T+\sigma^2\bone\preceq\bzero,\quad\bp\succeq\bzero.
  \end{split}
\end{equation}
where $\bdelta_k$ is the $k$-th row of $-(\bDelta^H)^{-1}$.
Note that problem (\ref{alternating_problem}) is equivalent to the problem (\ref{solvable_problem_SINR}); thus, it does not require any condition such as $L\ge K+Z$. Unfortunately, directly tackling (\ref{alternating_problem}) is still very difficult. This is because $\bLambda^{1/2}$ defined by (\ref{optimal_Lambda}) is non-convex in $x_j$. Also, the third constraint of (\ref{alternating_problem}) is non-convex in $x_j$ and $\eta$. Even if we assume other variables are fixed except $x_j$, the problem (\ref{alternating_problem}) becomes non-convex in $x_j$, which is very difficult to solve.

In the following, based on (\ref{alternating_problem}), we propose an alternating algorithm which is asymptotically optimal. Specifically,
we consider the asymptotic situation $\Ptot\rightarrow\infty$, which means that the total power of FJ and BS can be large. Before proposing an asymptotically optimal algorithm, in the following lemma, we first derive an important property of optimal $\{x_j\}$ when $\Ptot\rightarrow\infty$.
\begin{lemma}\label{lemma6}
When the condition of (\ref{existence_condition}) is satisfied, the optimal solution $\{x_j\}$ to the problem (\ref{alternating_problem}) must satisfy
$\lim_{\Ptot\rightarrow\infty} x_j\rightarrow\infty, j=1,\cdots,Z$.
\end{lemma}
\begin{IEEEproof}
See Appendix \ref{proof_lemma6}.
\end{IEEEproof}
When $\Ptot\rightarrow\infty$, it follows from Lemma \ref{lemma6} that
%the definition of $\bLambda^{1/2}$ and the fourth constraint in (\ref{alternating_problem}), $x_j^2$ goes to infinity and $\eta$ goes to $\max_k \{p_k\left(\sum_{j=Z+1}^{Z}\frac{|a_{kj}|^2}{\sigma^2}\right)\}$ as $\Ptot\rightarrow\infty$, which means
$\lim_{\Ptot\rightarrow\infty}\frac{x_j^2}{\sigma^2+x_j^2}=1$.
Using this asymptotic result in (\ref{alternating_problem}), it is possible to derive an asymptotic version of the alternating algorithm. Denoting $\btc:=[\|\bc_1\|^2,\cdots,\|\bc_K\|^2]^T$, we can write $\bp=[p_1,p_2,\cdots,p_K]^T$, where $p_k=\bdelta_k^H(\btc+\sigma^2\bone)$. Also, we write $\bSigma(\bGamma)=\bGamma^H\bGamma$, where $\bGamma$ can be determined by given $\btc$ and $\{x_j: j=1,\cdots,Z\}$. Then we propose an alternating algorithm to obtain $\btc$ and $\{x_j: j=1,\cdots,Z\}$.

\emph{Alternating Algorithm:}
\begin{itemize}
\item Initialize $\btc=\bzero$.
\item In each iteration:
    \begin{itemize}
    \item[-] Step 1: Given $\btc$, $\{x_j: j=1,\cdots,Z\}$ are updated by the following convex optimization problem:
    \begin{equation}\label{alternating_step1}
    \begin{split}
      \{x_j\}&=\arg_{\{x_j\}}\min_{\{x_j\ge 0\},\bGamma,\eta}\eta\quad\\
      \st\quad& \bG^H\bGamma^H=[\diag\{x_1,\cdots,x_{Z}\},\bzero]^T,\\
      &\tr\{\bGamma^H\bGamma\}\le\Ptot-\sum_{k=1}^K\bdelta_k^H(\btc+\sigma^2\bone),\\
      &\sum_{j=1}^{Z}\frac{|a_{kj}|^2}{x_j^2}\le\frac{\eta}{\bdelta_k^H(\btc+\sigma^2\bone)},\quad k=1,\cdots,K,\\
      &\left[\bb_1^H\bGamma^H\bGamma\bb_1,\cdots,\bb_K^H\bGamma^H\bGamma\bb_K\right]^T\preceq\btc.
    \end{split}
    \end{equation}

    \item[-] Step 2: Given $\{x_j: j=1,\cdots,Z\}$, $\btc$ is updated by the following convex optimization problem:
    \begin{equation}\label{alternating_step2}
    \begin{split}
      \btc=&\arg_{\btc}\min_{\btc,\bGamma,\eta}\eta\quad\\
      \st\quad& \bG^H\bGamma^H=[\bLambda^{1/2},\bzero]^T, \\ &\left[\bb_1^H\bGamma^H\bGamma\bb_1,\cdots,\bb_K^H\bGamma^H\bGamma\bb_K\right]^T\preceq\btc\\
      &\tr\{\bGamma^H\bGamma\}+\sum_{k=1}^K\bdelta_k^H(\btc+\sigma^2\bone)\le\Ptot,\\ &0\le\bdelta_k^H(\btc+\sigma^2\bone)\le\frac{\eta}{\sum_{j=1}^{Z}\frac{|a_{kj}|^2}{x_j^2}},k=1,\cdots,K.
    \end{split}
    \end{equation}
    \end{itemize}
\end{itemize}
Note that above alternating algorithm must converge to a critical point, since in each step the value of objective function is monotonically decreasing and the optimal value is bounded.
More importantly, the proposed alternating algorithm does not require any condition on the number of antennas at FJ, and thus, it can be applied to both $L\ge K+Z$ and $L<K+Z$.

Although the alternating algorithm gives a suboptimal solution to (\ref{solvable_problem_SINR}), we can prove that if  $L\ge K+Z$, the proposed alternating algorithm is asymptotically optimal as $\Ptot\rightarrow\infty$, which is given in the following Lemma:
\begin{lemma}\label{lemma3}
When $L\ge K+Z$ and the condition (\ref{existence_condition}) is satisfied, the proposed alternating algorithm is asymptotically optimal in the sense that as $\Ptot\rightarrow\infty$, its solution converges to the optimal solution.
\end{lemma}
\begin{IEEEproof}
See Appendix \ref{proof_lemma3}.
\end{IEEEproof}
From Lemma \ref{lemma3}, one knows that, when $L\ge K+Z$, the proposed alternating algorithm is asymptotically optimal in the sense of $\Ptot\rightarrow\infty$. Then a natural question arising is whether the proposed alternating algorithm is still asymptotically optimal in any sense when $L<K+Z$. In the following, we answer this question. Specifically, the answer is that, when $L<K+Z$, the proposed alternating algorithm is asymptotically optimal in the sense of $\bB\rightarrow\bzero$ and $\Ptot\rightarrow\infty$. Note that $\bB\rightarrow\bzero$ is an important asymptotic case for the following reason. As discussed before, when $L<K+Z$, the zero-forcing is not optimal, meaning that, when $L<K+Z$, the jamming signal must be received by the users with the optimal CJ. Thus, when $L<K+Z$, using CJ becomes more effective only when the channel $\bB$ from FJ to the users becomes weaker, i.e., $\bB\rightarrow\bzero$. On the other hand, when $L<K+Z$, if $\bB\rightarrow\infty$, using CJ is not an effective approach because the users will be significantly affected by the jamming signal. When $L<K+Z$, therefore, $\bB\rightarrow\bzero$ is an important asymptotic case where adopting the approach of CJ is justified and recommended.

In order to show the asymptotic optimality of the proposed alternating algorithm in the sense of $\bB\rightarrow\bzero$ and $\Ptot\rightarrow\infty$, we first study the extreme case that the channel $\bB$ is completely blocked, i.e., $\bB=\bzero$ and $\Ptot\rightarrow\infty$.
%From Lemma \ref{lemma3}, we know that the alternating algorithm converges to the optimal algorithm when $L\ge K+Z$. On the other hand, when $Z\le L<K+Z$, designing CJ may not be possible in general, since the DoF of the FJ is not enough , as we discussed before. However, even when $Z\le L<K+Z$, one might design CJ if the jamming generated by CJ does not significantly affect the users, which happens when the channel $\bB$ from the FJ to the users is weak. In the following, therefore, we consider an asymptotic case of $\bB\rightarrow\bzero$. We first study the extreme case that the channel between the FJ and legitimate users is completely blocked, i.e., $\bB=\bzero$, and give the asymptotically optimal solution when $\Ptot\rightarrow\infty$.
\begin{lemma}\label{lemma4}
If $\bB=\bzero$ and the condition of (\ref{existence_condition}) is satisfied, the asymptotically optimal solution to (\ref{solvable_problem_SINR}) when $\Ptot\rightarrow\infty$ can be obtained by the following convex optimization problem:
\begin{equation}\label{equ_lemma4}
  \begin{split}
    \min_{\bGamma,\{x_j\},\eta}\quad\eta\quad
      \st\quad&\bG^H\bGamma^H=[\bLambda^{1/2},\bzero]^T\bV^H,\quad \\ &\tr\{\bGamma^H\bGamma\}-\sigma^2\|(\bDelta^H)^{-1}\bone\|_1\le\Ptot,  \\
      &p_k\sum_{j=1}^{Z}\frac{|a_{kj}|^2}{x_j^2}\le\eta, \quad k=1,\cdots,K.\\
%      &\bb_k^H\bGamma^H=\bc_k^H,\quad k=1,\cdots,K.
      %&\bDelta^H\bp+[\|\bc_1\|^2,\cdots,\|\bc_K\|^2]^T+\sigma^2\bone\preceq\bzero,\quad\bp\succeq\bzero.
  \end{split}
\end{equation}
\end{lemma}
\begin{IEEEproof}
Substituting $\bB=\bzero$ into (\ref{alternating_problem}) and using the asymptotic result $\sigma^2+x_j^2\rightarrow x_j^2$, it is easy to see that $\{\bc_k=\bzero: k=1,\cdots,K\}$ is optimal. Then we obtain (\ref{equ_lemma4}), which is a convex optimization problem of $\bGamma$, $\{x_j\}$, and $\eta$.
\end{IEEEproof}

When $\bB=\bzero$, the result of Lemma \ref{lemma4} can be directly used. On the other hand, for the case of $\bB\rightarrow\bzero$ (but $\bB\neq\bzero$) which we are interested in, the result of Lemma \ref{lemma4} cannot be directly used since the interference to the legitimate users must be taken into account. The usefulness of Lemma \ref{lemma4} is that it can be used to prove an asymptotic optimality of the proposed alternating algorithm for the case $\bB\rightarrow\bzero$.

In the following lemma, we prove that the solution obtained by the proposed alternating algorithm converges to the asymptotically optimal solution in Lemma \ref{lemma4} when $\bB\rightarrow\bzero$.
\begin{lemma}\label{lemma5}
When $\bB\rightarrow\bzero$, the proposed alternating algorithm in (\ref{alternating_step1}) and (\ref{alternating_step2}) is asymptotically optimal in the sense that its solution converges to the optimal solution of Lemma \ref{lemma4} as $\Ptot\rightarrow\infty$.
\end{lemma}
\begin{IEEEproof}
See Appendix \ref{proof_lemma5}.
\end{IEEEproof}
From the results of Lemmas \ref{lemma3} and \ref{lemma5}, the proposed alternating algorithm can be considered as a very effective suboptimal method. Specifically, if $L\ge K+Z$ the performance of the proposed alternating algorithm converges to the optimal performance given by Theorem \ref{theorem1} as $\Ptot\rightarrow\infty$. Also, if $L<K+Z$ and the channel $\bB$ between FJ and legitimate users is weak, the performance of the proposed suboptimal algorithm converges to optimal performance given by Lemma \ref{lemma4} when $\Ptot\rightarrow\infty$. These results will be numerically confirmed in Section \ref{section_simulations}.

\subsection{Comparison with Existing CJ}
Most of the existing work on CJ, such as \cite{Liao2011,Li2011,Huang2011a,Huang2012}, did not consider multiple users or multiple data streams. Only recently, the design of CJ for multiple users with multiple streams has been studied in \cite{YangSubmitted,YangSubmitted2}. However, the problem of (\ref{solvable_problem}) and the obtained results are substantially different from those of the existing CJ methods such as \cite{YangSubmitted} and \cite{YangSubmitted2}. In \cite{YangSubmitted}, the problem of minimizing the CJ power was considered when multiple eavesdroppers existed. Since the cost function considered in \cite{YangSubmitted} is different from that of this paper, the CJ solution in \cite{YangSubmitted} is not comparable with the results in this paper. Also, the limitation of \cite{YangSubmitted} is that the obtained CJ power could be very high, which may not be practical. On the other hand, the problem in \cite{YangSubmitted2} is similar to problem (\ref{solvable_problem_SINR}) of this paper: the problem is to minimize the maximum achievable SINR at Eve subject to the CJ power is constrained. The differences between \cite{YangSubmitted2} and this paper are as follows: First, the optimization problems are different. In \cite{YangSubmitted2}, we considered to minimize the total power under the SINR constraints on legitimate users and eavesdropper, However, in (\ref{solvable_problem_SINR}) , joint design of the power allocation and CJ is considered to minimize a upper bound of the SINR at eavesdropper. Furthermore, we also give an equivalent formulation of our design problem in terms of the lower bound of the achievable secrecy rate. Second, the precoding matrix $\bW$ was assumed to be known in \cite{YangSubmitted2}, which means that no joint optimization was considered at all between the BS and the FJ. Since fully joint design of $\bW$ and CJ is analytically intractable in general, we consider a balanced problem in this paper to design the CJ and partial $\bW$. Furthermore, in \cite{YangSubmitted2} the CJ was simply made orthogonal to the users' channel without proving its optimal sense. On the other hand, in this paper, it is proved that such zero-forcing is optimal only when $L\ge K+Z$. We also consider the case $L<K+Z$ which is much more difficult than $L\ge K+Z$ and an asymptotic optimal algorithm is derived. Moreover, in \cite{YangSubmitted2} the SINR in the form of (\ref{normal_SINR}) was used in the objective function, rather than the upper bound of the SINR of (\ref{worst_case_SINR}), meaning that the results of \cite{YangSubmitted2} might be rather optimistic from the perspective of the users. If the upper bound of the SINR is used in \cite{YangSubmitted2}, it can be shown that the result of \cite{YangSubmitted2} is a special case of the result in this paper.
\begin{lemma}\label{lemma7}
When the upper bound of the SINR (\ref{worst_case_SINR}) is used in the problem \cite[Eq. (1)]{YangSubmitted2}, by replacing $\bR_0$ by $\bI$ in \cite{YangSubmitted2}, the solution given by \cite[Eq. (2) and Eq. (3)]{YangSubmitted2} is still valid, which can be written as the same forms of (\ref{optimal_Gamma}) and (\ref{optimal_Lambda}), in which $\bx$ is determined by:
\begin{equation}\label{temp_lemma7}
    \begin{split}
    \bx&=\arg\min_{\bzero_{Z\times 1}\prec \bx\preceq\frac{1}{\sigma^2}\bone_{Z\times 1},\eta}\eta\quad\\
    \st&\sum_{j=1}^{Z}\phi_j{x_j}^{-1}\le\sigma^2\sum_{j=1}^{Z}\phi_j+\Pmax,\\
    &\sum_{j=1}^{Z}|\sqrt{p_k}a_{kj}|^2 x_j\le\eta,k=1,\cdots,K,
\end{split}
\end{equation}
where $\Pmax$ is the CJ power constraint.
\end{lemma}
\begin{IEEEproof}
From the definition of $\tilde{a}_{kj}$ in \cite{YangSubmitted2}, one can see that $\tilde{a}_{kj}=\sqrt{p_k}a_{kj}$. Thus, when $\bp$ is given, the problem (\ref{optimal_lambda}) becomes equivalent to \cite[Eq. (3)]{YangSubmitted2} by denoting $(\Ptot-\|\bp\|_1)$ as $\Pmax$ and $(x_j^{-1}-1)$ as $\lambda_j$.
\end{IEEEproof}
Comparing (\ref{temp_lemma7}) to (\ref{optimal_lambda}), one can see that (\ref{optimal_lambda}) is more general than (\ref{temp_lemma7}) in the sense that the individual power $p_k$ is optimized in (\ref{optimal_lambda}) along with (\ref{optimal_p}), whereas the individual power is assumed to be simply given in (\ref{temp_lemma7}). If the power allocation vector $\bp$ in (\ref{optimal_lambda}) is assumed to be given without optimization, then (\ref{optimal_lambda}) reduces to (\ref{temp_lemma7}). Therefore, the result of \cite{YangSubmitted2} can be seen as a special case of the result of this paper.
%According to Corollary \ref{lemma7}, since the linear precoding matrix is determined when $\bp$ is given, the problem in this paper reduces to the problem in \cite{YangSubmitted2} when the SINR of (\ref{worst_case_SINR}) is applied in \cite{YangSubmitted2}. In other words, the solution of Theorem \ref{theorem1} can be seen as a generalization of the previous work \cite{YangSubmitted2} since partial joint designing of linear precoding matrix and CJ is discussed in this paper.

\section{Simulations}\label{section_simulations}
In this section, we investigate the performance of the proposed algorithms numerically. We set the noise power $\sigma^2=-10$ dBm. The channel matrices $\bH$, $\bG$, $\bB$, and $\bF$ are generated according to Rayleigh fading such that the power gain of each element of the matrices is $0$ dB. For the BS, we assume the normalized linear precoding vectors are obtained by the very well-known channel inversion algorithm \cite{Peel2005}, i.e., $\bu_k=\frac{\tilde{\bu}_k}{\|\tilde{\bu}_k\|}$ where $\tilde{\bu}_k$ is the $k$-th column of $\bF(\bF^H\bF)^{-1}$. Monte Carlo experiments consisting of $10^3$ independent trials are performed to obtain the average results. Note that the complexity of optimal solution mainly depends on (i) the computation of $\bSigma_{\textrm{opt}}$, which is $\mathcal{O}(L^3)$, and (ii) solving $\bx$ by convex optimization problem with $Z$ variables, which is about $\mathcal{O}(Z^3)$. For the proposed suboptimal algorithm, the computational complexity of the iterative algorithm mainly depends on (i) the number of iterations, which is around $5-15$ in our examples, and (ii) the complexity of solving two convex optimization problems, each with $L^2/2+Z$ variables in an single alternating iteration. So the computational complexity is about $\mathcal{O}(L^6)$, which is about $10^2$ times of the optimal algorithm in our numerical examples.

\begin{figure}
\centering\includegraphics[width=0.37\textwidth]{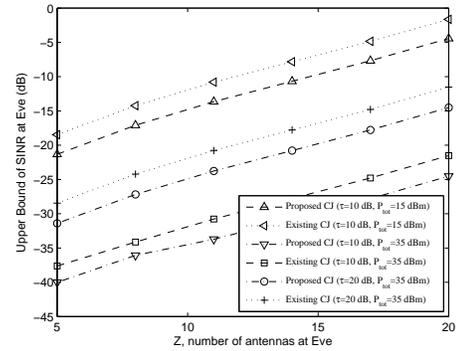}
\caption{Upper bound of the SINR at Eve versus the number of antennas at Eve. Proposed optimal solution and the existing method \cite{YangSubmitted2}.}\label{fig1}
\end{figure}
\begin{figure}
\centering\includegraphics[width=0.37\textwidth]{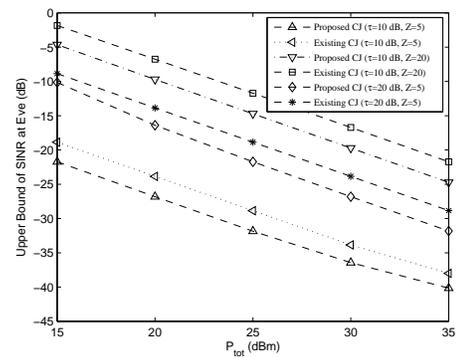}
\caption{Upper bound of the SINR at Eve versus the total power of FJ and BS. Proposed optimal solution and the existing method \cite{YangSubmitted2}.}\label{fig2}
\end{figure}
\begin{figure}
\centering\includegraphics[width=0.37\textwidth]{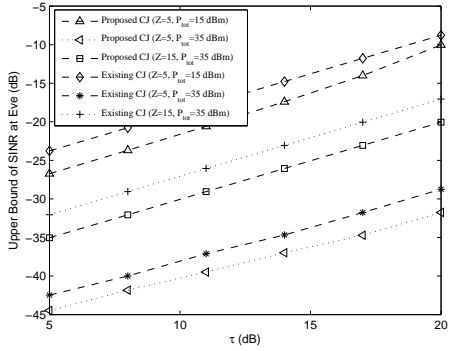}
\caption{Upper bound of the SINR at Eve versus the QoS threshold for users. Proposed optimal solution and the existing method \cite{YangSubmitted2}.}\label{fig3}
\end{figure}
The optimal algorithm when $L\ge K+Z$ is investigated in the first three examples and the minimized upper bound of the SINR at Eve by (\ref{solvable_problem_SINR}) is demonstrated. We set $N=20$, $K=10$, and $L=35$ as default values, and change the values of $\Ptot$, $\tau$, and $Z$ in different examples. For comparison, we also included the existing CJ in \cite{YangSubmitted2} using the upper bound of the SINR of (\ref{worst_case_SINR}) as secure metric, which is also given in (\ref{temp_lemma7}). For (\ref{temp_lemma7}), we assume half power of $\Ptot$ is allocated to the BS and the other half of $\Ptot$ is allocated to the FJ, i.e., $\Pmax=\frac{1}{2}\Ptot$. In the first example, the number $Z$ of antennas at Eve, is varied from $5$ to $20$ and the upper bound of the SINR defined by (\ref{worst_case_SINR}) is plotted in Fig. \ref{fig1}. From the figure, one can see that the upper bound of the SINR increases by nearly $10$ dB when $Z$ increases from $5$ to $20$. Also, the upper bound of the SINR at Eve is lower when $\Ptot$ is larger or $\tau$ is lower.
In the second example, we vary $\Ptot$ from $15$ dBm to $35$ dBm, which is shown in Fig. \ref{fig2} for different cases of $Z$ and $\tau$. According to the figure, increasing $\Ptot$ is an effective way to reduce the upper bound of the SINR at Eve, enhancing the security of the network. We can also see from Fig. \ref{fig2} that upper bound of the SINR increases if more antennas are employed at Eve.
In the third example, the QoS threshold, $\tau$, for users is changed from $5$ dB to $20$ dB. The corresponding upper bound of the SINR at Eve is shown in Fig. \ref{fig3}. It is shown that by increasing the QoS for users, the upper bound of the SINR at Eve increases as well. This is because the power of data streams received by Eve increases and also the capability of CJ is limited since the power for CJ is reduced.
Note that in all the three examples, the proposed optimal CJ is always better than the existing CJ of (\ref{temp_lemma7}), because the optimal power allocation between the BS and FJ is jointly designed with CJ in the proposed algorithm.

\begin{figure}
\centering\includegraphics[width=0.37\textwidth]{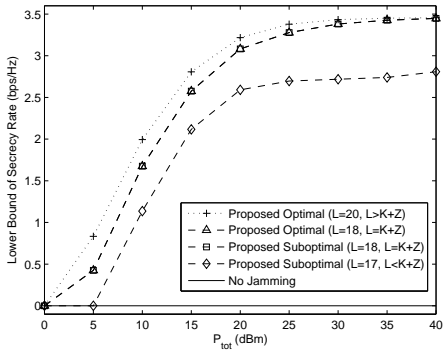}
\caption{Lower bound of secrecy rate versus the total power of FJ and BS.}\label{fig4}
\end{figure}
\begin{figure}
\centering\includegraphics[width=0.37\textwidth]{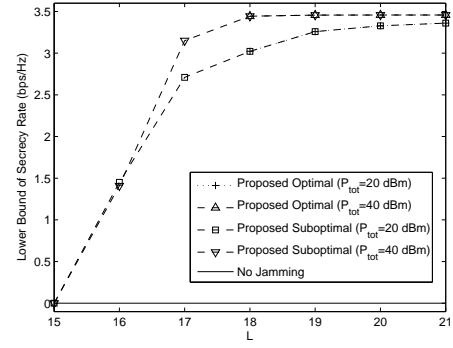}
\caption{Lower bound of secrecy rate versus the number of antennas at FJ. $K+Z=18$.}\label{fig5}
\end{figure}
\begin{figure}
\centering\includegraphics[width=0.37\textwidth]{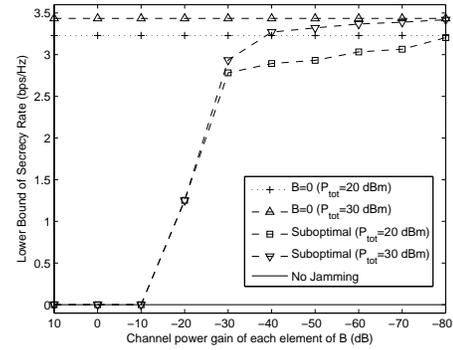}
\caption{Lower bound of secrecy rate versus the channel $\bB$ between FJ and legitimate users.}\label{fig6}
\end{figure}
\begin{figure}
\centering\includegraphics[width=0.37\textwidth]{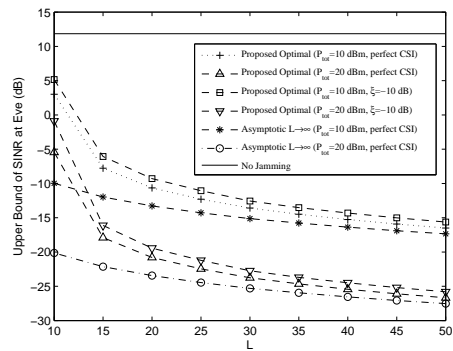}
\caption{Imperfect CSI and asymptotic performance of $L\rightarrow\infty$.}\label{fig7}
\end{figure}
In the next three examples, we investigate the proposed suboptimal alternating algorithm and the maximum of the lower bound of the secrecy rate by (\ref{solvable_problem}) is demonstrated. Each element of channel $\bB$ is generated such that the power gain of each element of $\bB$ is  $-30$ dB. We set $N=10$, $K=3$, $Z=15$, and $\tau=10$ dB as default values, and change $L$ and $\Ptot$ in each example. First, we change $\Ptot$ for different values of $L$, which is shown in Fig. \ref{fig4}. In the figure, the optimal algorithm is plotted for $L=20>K+Z$ and $L=K+Z=18$ for comparison, and the suboptimal alternating algorithm is plotted for $L=K+Z=18$ and $L=17<K+Z$. We also included the lower bound of the secrecy rate when there is no CJ. According to Fig. \ref{fig4}, the lower bound of the secrecy rate is increasing when $\Ptot$ is increasing. From the case $L=18$, one can see that the proposed suboptimal alternating algorithm converges to the optimal algorithm when $\Ptot$ is large, e.g., larger than $5$ dBm in our example, the performance of two algorithms is very close to each other.
Next, we change $L$ from $15$ to $21$ and fix $\Ptot$ equals to $20$ dBm or $40$ dBm. The resulting lower bound of the secrecy rate is shown in Fig. \ref{fig5}. In the figure, the optimal algorithm is shown only when $L\ge K+Z$, whose performance is essentially the same as the performance of the proposed alternating algorithm. One can also see that the effect of transmitting CJ is severely limited by the number $L$. For example, when $L=15$ even if we set $\Ptot=40$ dBm, the lower bound of the secrecy rate is almost the same as the case when no CJ is transmitted, which is close to $0$ bps/Hz. Thus, the CJ is not very useful when $L$ is much lower than $K+Z$.
Finally, letting $K=3$, $Z=15$, and $L=17<K+Z$, we generate $\bB$ according to Rayleigh fading with different power gain, from $10$ dB to $-80$ dB. The performance of the suboptimal algorithm is plotted compared with the asymptotic case $\bB=\bzero$ in Fig. \ref{fig6}. One can see that as $\bB\rightarrow\bzero$, the proposed suboptimal algorithm asymptotically converge to the optimal performance when $\bB=\bzero$. This means even if $L<K+Z$, the proposed suboptimal algorithm can be very effective when the channel between the FJ and legitimate users is very weak.

In the last example, we considered the Eve's channels $\bG$ and $\bB$ are perturbed by a Gaussian noise with variance $\xi^2=-10$ dB, i.e. $\hat{\bG}=\bG+\bDelta_G$, $\hat{\bB}=\bB+\bDelta_B$, where $\bG\sim\mathcal{CN}(\bzero,\bI)$, $\bB\sim\mathcal{CN}(\bzero,\bI)$, $\bDelta_G\sim\mathcal{CN}(\bzero,0.1\bI)$, and $\bDelta_B\sim\mathcal{CN}(\bzero,0.1\bI)$. From the results in Fig. \ref{fig7}, one can see that the performance of the proposed optimal CJ scheme deteriorates with imperfect CSI. However, the performance is still much better than the case of no jamming. We also include the performance limit for $L\rightarrow\infty$. When $L\rightarrow\infty$, the channels of $\{\bb_k\}$ and $\{\bg_j\}$ tend to be uncorrelated. Thus, we have $\bGamma^H_{\textrm{opt}}\approx\left(\frac{\sqrt{x_1^{-1}-\sigma^2}}{|\bg_1|^2}\bg_1,\cdots,\frac{\sqrt{x_Z^{-1}-\sigma^2}}{|\bg_Z|^2}\bg_Z\right)$ and $\phi_j$ can be replaced by $\frac{1}{|\bg_j|^2}$ since the matrix $\left[\bG^H\bG-\bG^H\bB\left(\bB^H\bB\right)^{-1}\bB^H\bG\right]^{-1}$ reduces to $\diag\{\frac{1}{|\bg_1|^2},\cdots,\frac{1}{|\bg_Z|^2}\}$ when $L\rightarrow\infty$. However, asymptotically $\frac{1}{\|\bg_j\|^2}\rightarrow 0$ as $L\rightarrow\infty$. Thus, $\eta\rightarrow 0$. From Fig. \ref{fig7}, one can see that, as $L$ increases, the performance of the proposed optimal algorithm gets close to the performance limit of $L\rightarrow\infty$.

\section{Conclusion}
We have proposed optimal and suboptimal algorithms for joint design of the power allocation between different users at BS and the CJ at the FJ to maximize a lower bound of secrecy rate. Compared to existing works, our problem is more general in the sense that joint optimizations are carried out. We demonstrated the proposed CJ could effectively interfere Eve to help the BS communicate confidentially with the legitimate users. In particular, in order to make the CJ strategy effective, it is important to employ enough number of antennas at the FJ. Moreover, increasing the total power and choosing relatively small $\tau$ could also enhance the security level. Finally, if the channel $\bB$ is weak, the CJ could also be effective even if $L<K+Z$.

\appendices
\numberwithin{equation}{section}

\section{}\label{proof_remark}
Let $s_k(t)$ denote the $k$-th stream with $|s_k(t)|^2=1$; then the received signal of the $k$-th stream at Eve can be written as $\br(t)=\sum_{k=1}^K\sqrt{p_k}\bH^H\bu_k s_k(t)+\bn(t)+\bG^H\bJ(t)$. Note that since $\bJ(t)$ is Gaussian, the term $\bn(t)+\bG^H\bJ(t)$ can be seen as a colored Gaussian noise with covariance matrix $\sigma^2\bI+\bG^H\bSigma\bG$.
Denoting $\ba_k=\bH^H\bu_k$, the ML detection at Eve can be written as
\begin{equation}\label{temp_remark_ML}
\begin{split}
  &\max_{\{s_k(t): k=1,\cdots,K\}}\frac{1}{\textrm{det}(\pi\left(\sigma^2\bI+\bG^H\bSigma\bG\right))}\\
  &\cdot e^{-\tr\left[(\br(t)-\sum_{k=1}^K\sqrt{p_k}\ba_k s_k(t))^H\bSigma^{-1}(\br(t)-\sum_{k=1}^K\sqrt{p_k}\ba_k s_k(t))\right]}.
\end{split}
\end{equation}
If we consider the upper bound of the SINR, the received signal at Eve can be written as $\br_k(t)=\sqrt{p_k}\bH^H\bu_k s_k(t)+\bn(t)+\bG^H\bJ(t)$. Then the ML detection at the eavesdropper for the $k$-th stream can be written as
\begin{equation}\label{temp_remark_WC_ML}
\begin{split}
  &\max_{s_k(t)}\frac{1}{\textrm{det}(\pi\left(\sigma^2\bI+\bG^H\bSigma\bG\right))}\\
  &\cdot e^{-\tr\left[(\br_k(t)-\sqrt{p_k}\ba_k s_k(t))^H\bSigma^{-1}(\br_k(t)-\sqrt{p_k}\ba_k s_k(t))\right]}.
\end{split}
\end{equation}
Let $P_s^{\textrm{U-ML}}$ denotes the SER of (\ref{temp_remark_WC_ML}). Then it is obvious that $P_s^{\textrm{U-ML}}\le P_s^{\textrm{ML}}$.

Next, we prove that $P_s^{\textrm{U-SINR}}=P_s^{\textrm{U-ML}}$. Note that (\ref{temp_remark_WC_ML}) is equivalent to
\begin{equation}\label{}
\begin{split}
  \min_{s_k(t)}&\|\left(\sigma^2\bI+\bG^H\bSigma\bG\right)^{-1/2}\br_k(t)\\
  &-\left(\sigma^2\bI+\bG^H\bSigma\bG\right)^{-1/2}\sqrt{p_k}\ba_k s_k(t)\|^2.
\end{split}
\end{equation}
Let $\hat{\br}_k(t):=\left(\sigma^2\bI+\bG^H\bSigma\bG\right)^{-1/2}\br_k(t)$ and $\hat{\ba}_k:=\sqrt{p_k}\left(\sigma^2\bI+\bG^H\bSigma\bG\right)^{-1/2}\ba_k$; then the ML estimate is
$\hat{s}_k(t)=\frac{\hat{\ba}_k^H}{\|\hat{\ba}_k\|^2}\hat{\br}_k(t)$.
We can show that
$\hat{s}_k(t)\sim \mathcal{CN}\left(s_k(t),\frac{1}{\|\tilde{\ba}_k\|^2}\right)$,
and the SINR which is actually SNR, is given by
$\|\tilde{\ba}_k\|^2=p_k\ba_k^H\left(\sigma^2\bI+\bG^H\bSigma\bG\right)^{-1}\ba_k$.

On the other hand, by maximizing the upper bound of the SINR, we get
\begin{equation}\label{}
    \begin{split}
    &p_k\bu_k^H\bH^H\left(     p_k\bH^H\bu_k\bu_k^H\bH+\sigma^2\bI+\bG^H\bSigma\bG\right)^{-1}\bH\bu_k\\
%    &=p_k\ba_k^H\left(p_k\ba_k\ba_k^H+\sigma^2\bI+\bG^H\bSigma\bG\right)^{-1}\ba_k\\
   % &=p_k\ba_k^H\left(\sigma^2\bI+\bG^H\bSigma\bG\right)^{-1}\ba_k
   % -\frac{p_k\ba_k^H\left(\sigma^2\bI+\bG^H\bSigma\bG\right)^{-1}\ba_kp_k\ba_k^H\left(\sigma^2\bI+\bG^H\bSigma\bG\right)^{-1}\ba_k}{1+p_k\ba_k^H\left(\sigma^2\bI+\bG^H\bSigma\bG\right)^{-1}\ba_k}\\
    &=\frac{p_k\ba_k^H\left(\sigma^2\bI+\bG^H\bSigma\bG\right)^{-1}\ba_k}{1+p_k\ba_k^H\left(\sigma^2\bI+\bG^H\bSigma\bG\right)^{-1}\ba_k}.
    \end{split}
\end{equation}
Then
$\SINR^U_{e,k}(p_k,\bSigma)
    =p_k\ba_k^H\left(\sigma^2\bI+\bG^H\bSigma\bG\right)^{-1}\ba_k$.
Thus, ML decoding is equivalent to optimal receive beamforming when the upper bound of the SINR is used, which means $P_s^{\textrm{U-SINR}}=P_s^{\textrm{U-ML}}\le P_s^{\textrm{ML}}$.

\section{Proof of Lemma \ref{lemma1}}\label{proof_lemma1}
The existence condition of (\ref{solvable_problem}) is equivalent to the existence condition for $\bp\succeq\bzero$ and $\bSigma$ that satisfies both $\|\bp\|_1+\tr(\bSigma)\le\Ptot$ and $\SINR_k(\bp,\bSigma)\ge\tau$. Note that $\|\bp\|_1\le\Ptot-\tr(\bSigma)\le\Ptot$ and $\tau\le\SINR_k(\bp,\bSigma)\le\SINR_k(\bp,\bzero)$. Thus, the existence condition for $\bp$ is equivalent to the condition that satisfies $\|\bp\|_1\le\Ptot$ and $\tau\le\SINR_k(\bp,\bzero)$, which is the traditional non-secure problem for linear precoding design \cite{Schubert2004,Schubert2005}. The constraints $\tau\le\SINR_k(\bp,\bzero)$ for $k=1,\cdots,K$ can be written as $\bDelta^H\bp+\sigma^2\bone\preceq\bzero$. Thus, we can prove that $\|\bp\|_1$ is minimized when the equality in $\bDelta^H\bp+\sigma^2\bone\preceq\bzero$ holds. Finally, the existence condition is $\|-\sigma^2\left(\bDelta^H\right)^{-1}\bone\|_1\le\Ptot$ and $\|-\sigma^2\left(\bDelta^H\right)^{-1}\bone\|_1\succeq\bzero$.
\section{Proof of Lemma \ref{lemma2}}\label{proof_lemma2}
We prove Lemma \ref{lemma2} by two steps. First, assuming $\bSigma=\bGamma^H\bGamma$, we reformulate the equivalent problem (\ref{solvable_problem_SINR}) so that $\bb_k^H\bGamma^H$ are denoted as new variables. Next, we prove $\|\bb_k^H\bGamma^H\|^2=0$, for all $k=1,\cdots,K$, are satisfied for the optimal solution, which implies $\bb_k^H\bSigma\bb_k=0$, for all $k=1,\cdots,K$. Thus, the property $\bB^H\bSigma=\bzero$ holds.

\subsection{Step 1: Reformulation of (\ref{solvable_problem_SINR})}
we can introduce another variable $\eta$ as $\eta=\max_k\{\SINR^U_{e,k}\}$ and add the following new constraints:
$    p_k\bu_k^H\bH^H\left(     p_k\bH^H\bu_k\bu_k^H\bH+\sigma^2\bI+\bG^H\bSigma\bG\right)^{-1}\bH\bu_k\le\frac{\eta}{1+\eta}$.
Note that the CJ is only determined by $\bG^H\bSigma\bG$ which is a $Z\times Z$ matrix; thus, the rank of $\bSigma$ equals to $Z$, which is smaller than $L$.
Using $\ba_k=\bH^H\bu_k$, $\bSigma=\bGamma^H\bGamma$, and $\bQ=\bGamma\bG$, from the result in Appendix \ref{proof_remark}, the SINR constraint at Eve is equivalent to $p_k\ba_k^H\left(\bG^H\bSigma\bG+\sigma^2\bI\right)^{-1}\ba_k\le\eta$, which can be written as $\frac{p_k}{\sigma^2}\left[\ba_k^H\ba_k-\ba_k^H\bQ^H\left(\bQ\bQ^H+\sigma^2\bI\right)^{-1}\bQ\ba_k\right]\le\eta$.
%\begin{equation}\label{}
%\begin{split}
%    &p_k\ba_k^H\left(\bG^H\bSigma\bG+\sigma^2\bI\right)^{-1}\ba_k\le\eta.\\
%    \Leftrightarrow\quad&\frac{p_k}{\sigma^2}\left[\ba_k^H\ba_k-\ba_k^H\bQ^H\left(\bQ\bQ^H+\sigma^2\bI\right)^{-1}\bQ\ba_k\right]\le\eta.
%\end{split}
%\end{equation}
We denote the eigenvalue decomposition of $\bQ\bQ^H$ as $\bQ\bQ^H=\bV\bLambda\bV^H$, then
\begin{equation}\label{}
\begin{split}
    &\frac{p_k}{\sigma^2}\ba_k^H\left[\bI-\bLambda^{1/2}\left(\sigma^2\bI+\bLambda\right)^{-1}\bLambda^{1/2}\right]\ba_k\\
    &= \sum_{j=1}^{Z}\left(1-\frac{\lambda_j}{\sigma^2+\lambda_j}\right)\frac{p_k|a_{kj}|^2}{\sigma^2}
    =\sum_{j=1}^{Z}\frac{p_k|a_{kj}|^2}{\sigma^2+\lambda_j},
\end{split}
\end{equation}
where $\lambda_j$ is the $j$-th eigenvalue of $\bQ\bQ^H$.
Next, we consider the QoS constraints at the users:
\begin{equation}\label{}
\begin{split}\small
    &\frac{p_k|\bof_k^H\bu_k|^2}{\sum_{i\neq k}p_i|\bof_k^H\bu_i|^2+\bb_k^H\bSigma\bb_k+\sigma^2}\ge\tau\\
    &\Leftrightarrow\quad p_k\frac{|\bof_k^H\bu_k|^2}{\tau}\ge\sum_{i\neq k}p_i|\bof_k^H\bu_i|^2+\|\bb_k^H\bGamma^H\|^2+\sigma^2\Leftrightarrow\quad\\
    &  \left[|\bof_k^H\bu_1|^2,\cdots,|\bof_k^H\bu_{k-1}|^2, -\frac{|\bof_k^H\bu_k|^2}{\tau},|\bof_k^H\bu_{k+1}|^2,\cdots,|\bof_k^H\bu_K|^2\right]\bp\\
    &+\|\bb_k^H\bGamma^H\|^2+\sigma^2\le 0.
\end{split}
\end{equation}
Using the definition of $\bDelta\in\mathbb{C}^{K\times K}$, we can write all users' SINR constraints together.
Denoting $\bc_k=\bGamma\bb_k$, the design problem can be written as
\begin{equation}\label{}
\begin{split}
    \min_{\{\bc_k\},\bp,\bGamma,\eta}&\quad \eta\quad\\
    \st &\quad\bG^H\bGamma^H=\left(\bV\bLambda^{1/2}\right)^H,\quad \lambda_j\ge 0,\quad j=1,\cdots,Z,\\
    &\|\bGamma\|^2+\|\bp\|_1\le\Ptot,\quad \bb_k^H\bGamma^H=\bc_k^H, \quad k=1,\cdots, K,\\
    & p_k\sum_{j=1}^{Z}\frac{|a_{kj}|^2}{\sigma^2+\lambda_j}\le\eta, \quad k=1,\cdots,K,\\
    &\bDelta^H\bp+\left(\|\bc_1\|^2,\cdots,\|\bc_K\|^2\right)^H+\sigma^2\bone\preceq\bzero,\quad\bp\succeq\bzero.
\end{split}
\end{equation}
Note that for any orthogonal matrix $\tilde{\bV}$, we always have $\|\bGamma^H\tilde{\bV}\|^2=\|\bGamma\|^2$ and $\|\bc_k\|^2=\|\bb_k^H\bGamma^H\|^2=\|\bb_k^H\bGamma^H\tilde{\bV}\|^2$. Thus, we can simply remove $\bV$ by replacing $\bGamma^H\bV$ by $\bGamma^H$.

\subsection{Step 2: Proving $\|\bc_k\|^2=0$ for $k=1,\cdots,K$}
We first assume $\bp$ is given. Denoting $\bC=[\bc_1,\cdots,\bc_K]$, we have
\begin{equation}\label{}
\begin{split}
    \min_{\{\bc_k\},\bGamma,\eta}\quad\eta
    \quad\st&\quad\bG^H\bGamma^H=\bLambda^{1/2},\quad
    \tr\{\bGamma^H\bGamma\}\le\Ptot-\|\bp\|_1,\\
    &\quad\bB^H\bGamma^H=\bC^H,\quad \lambda_j\ge 0,\quad j=1,\cdots,Z,\\
    &\quad\sum_{j=1}^{Z}\frac{|a_{kj}|^2}{\sigma^2+\lambda_j}\le\frac{\eta}{p_k},\quad k=1,\cdots,K.
\end{split}
\end{equation}
Note that the last constraint is only related to $\lambda_j$, which are only used to determine $\bLambda$. Thus, we can first fix $\lambda_j$; so $\bGamma$ can be obtained as a function of $\lambda_j$ as follows:
\begin{equation}\label{}
    \min_{\bGamma}\quad\tr\{\bGamma^H\bGamma\}\quad\st\quad\left[
                                                             \begin{array}{c}
                                                               \bG^H \\
                                                               \bB^H \\
                                                             \end{array}
                                                           \right]\bGamma^H=\left[
                                                             \begin{array}{c}
                                                               \bLambda^{1/2} \\
                                                               \bC^H \\
                                                             \end{array}
                                                           \right].
\end{equation}
The solution to the above problem exists and has the following closed form:
\begin{equation}\label{}
    \bGamma^H=\left[
                \begin{array}{cc}
                  \bG & \bB \\
                \end{array}
              \right]\left[
                       \begin{array}{cc}
                         \bG^H\bG & \bG^H\bB \\
                         \bB^H\bG & \bB^H\bB \\
                       \end{array}
                     \right]^{-1}\left[
                                                             \begin{array}{c}
                                                               \bLambda^{1/2} \\
                                                               \bC^H \\
                                                             \end{array}
                                                           \right].
\end{equation}
Then we have
%\begin{equation}\label{}
%\begin{split}
 $   \tr\{\bGamma^H\bGamma\}%&=\tr\left\{\left[
%                \begin{array}{cc}
%                  \bLambda^{1/2} & \bC \\
%                \end{array}
%              \right]\left[
%                       \begin{array}{cc}
%                         \bPhi_{11} & \bPhi_{12} \\
%                         \bPhi_{21} & \bPhi_{22} \\
%                       \end{array}
%                     \right]\left[
%                                                             \begin{array}{c}
%                                                               \bLambda^{1/2} \\
%                                                               \bC^H \\
%                                                             \end{array}
%                                                           \right]\right\}\\
                    % &=\tr\{\bLambda^{1/2}\bPhi_{11}\bLambda^{1/2}\}+\tr\{\bC\bPhi_{22}\bC^H\}\\
                     =\sum_{j=1}^{Z}\phi_j\lambda_j+\|\bPhi_{22}^{1/2}\bC\|^2,
                     $
%\end{split}
%\end{equation}
where $\phi_j$ is the $j$-th diagonal element of $\bPhi_{11}$,
$    \bPhi_{11}:=\left\{\bG^H\left[\bI-\bB\left(\bB^H\bB\right)^{-1}\bB^H\right]\bG\right\}^{-1}$, and
$    \bPhi_{22}:=\left\{\bB^H\left[\bI-\bG\left(\bG^H\bG\right)^{-1}\bG^H\right]\bB\right\}^{-1}$.
%\end{split}
%\end{equation}
The variable $\bGamma$ can be replaced so that the residual variables are $\bC$, $\eta$, $\bp$, and $\lambda_j$:
\begin{equation}\label{non_relaxed_c}
\begin{split}
    &\min_{\bp\succeq\bzero,\lambda_j\ge 0,\bC}\quad\eta\quad\\
    \st&\quad p_k\sum_{j=1}^{Z}\frac{|a_{kj}|^2}{\sigma^2+\lambda_j}\le\eta,\quad k=1,\cdots,K,\\
    &\sum_{j=1}^{Z}\phi_j\lambda_j+\|\bPhi_{22}^{1/2}\bC\|^2+\|\bp\|_1\le\Ptot,\\
    &\bDelta^H\bp+\left(\|\bc_1\|^2,\cdots,\|\bc_K\|^2\right)^H+\sigma^2\bone\preceq\bzero.
\end{split}
\end{equation}
Based on the above problem, we can prove that $\|\bc_k\|^2=0$ as follows: First of all, relax the constraint $\bDelta^H\bp+\left(\|\bc_1\|^2,\cdots,\|\bc_K\|^2\right)^H+\sigma^2\bone\preceq\bzero$ to $\bDelta^H\bp+\sigma^2\bone\preceq\bzero$, then it is easy to see that the optimal variables $\lambda_j$, $\bC$, and $\bp$ must satisfy $\bPhi_{22}^{1/2}\bC=\bzero$ in the following relaxed problem:
\begin{equation}\label{relaxed_c}
\begin{split}
    \min_{\bp\succeq\bzero,\lambda_j\ge 0,\bC}&\quad\eta\quad\\
    \st&\quad p_k\sum_{j=1}^{Z}\frac{|a_{kj}|^2}{\sigma^2+\lambda_j}\le\eta,\quad k=1,\cdots,K,\\
    &\quad\bDelta^H\bp+\sigma^2\bone\preceq\bzero\\
    &\sum_{j=1}^{Z}\phi_j\lambda_j+\|\bPhi_{22}^{1/2}\bC\|^2+\|\bp\|_1\le\Ptot.
\end{split}
\end{equation}
Let $\bC'=(\bc_1',\cdots,\bc_K')$ and $\bp'$ be the optimal solution to (\ref{non_relaxed_c}); then one can see that $\bC'$ can be any matrix that satisfies $\bPhi_{22}^{1/2}\bC'=\bzero$ since the optimal value does not change once $\bPhi_{22}^{1/2}\bC'=\bzero$ is satisfied. Furthermore, if the optimal $\bp'$ for (\ref{non_relaxed_c}) also satisfies $\bDelta^H\bp'+\left(\|\bc_1'\|^2,\cdots,\|\bc_K'\|^2\right)^H+\sigma^2\bone\preceq\bzero$, the optimal solution to the relaxed problem (\ref{relaxed_c}) falls into the feasible set of the problem (\ref{non_relaxed_c}). Obviously, $\{\bc_k'=\bzero: k=1,\cdots,K\}$, which satisfies $\bPhi_{22}^{1/2}\bC'=\bzero$, is the optimal solution to the relaxed problem (\ref{non_relaxed_c}); and thus, they must be the optimal solution to (\ref{relaxed_c}). Therefore, we have $\{\|\bc_k\|^2=0: k=1,\cdots,K\}$, which implies that $\bB^H\bSigma=\bzero$.

\section{Proof of Theorem \ref{theorem1}}\label{proof_theorem1}
From (\ref{relaxed_c}) with $\bC=\bzero$, we can get the following non-convex optimization problem
\begin{equation}\label{C1}
\begin{split}
    \min_{\bp\succeq\bzero,\{\lambda_j\}}\eta&\quad
    \st\quad p_k\sum_{j=1}^{Z}\frac{|a_{kj}|^2}{\sigma^2+\lambda_j}\le\eta,k=1,\cdots,K,\\
    &\sum_{j=1}^{Z}\phi_j\lambda_j+\|\bp\|_1\le\Ptot,
    \bDelta^H\bp+\sigma^2\bone\preceq\bzero.
\end{split}
\end{equation}
By solving the above problem, the optimal power allocation $\bp$ can be obtained and the optimal CJ $\bSigma$ can be computed as $\bSigma=\bGamma^H\bGamma$, where
$    \bGamma^H=\left[
                \begin{array}{cc}
                  \bG & \bB \\
                \end{array}
              \right]\left[
                       \begin{array}{cc}
                         \bG^H\bG & \bG^H\bB \\
                         \bB^H\bG & \bB^H\bB \\
                       \end{array}
                     \right]^{-1}\left[
                                                             \begin{array}{c}
                                                               \bLambda^{1/2} \\
                                                               \bzero \\
                                                             \end{array}
                                                           \right]$,
in which $\bLambda^{1/2}=\diag\{\sqrt{\lambda_1},\cdots,\sqrt{\lambda_{Z}}\}$ and
$\lambda_j$ is the $j$-th eigenvalue of $\bQ\bQ^H$. Using new variables $x_j=\frac{1}{\sigma^2+\lambda_j}$ and $y_k=\frac{1}{p_k}$, the above problem turns to
\begin{equation}\label{C4}
    \begin{split}
    &\min_{\{y_k>0\},\{0<x_j\le 1\}}\eta\quad\\
    \st&\quad \sum_{j=1}^{Z} |a_{kj}|^2 x_j-\eta y_k\le 0,\\
    &\quad \sum_{j=1}^{Z}\phi_j(\frac{1}{x_j}-\sigma^2)+\sum_{k=1}^K \frac{1}{y_k}\le\Ptot,\\
    &\quad y_k\le\frac{|\bof_k^H\bu_k|^2}{\tau\sigma^2+\tau\sum_{i=1, i\neq k}^K\frac{|\bof_k^H\bu_i|^2}{y_i}},\quad k=1,\cdots,K.
\end{split}
\end{equation}
Note that the above problem is a non-convex optimization problem because the last constraint is non-convex. However, we can prove that the equality of the last constraint must hold, which makes it possible to reformulate the non-convex optimization problem to a convex optimization problem.

First, we prove that if $\{x_1,\cdots,x_Z, y_1,\cdots,y_K\}$ is a feasible point of (\ref{C4}) and the last constraint is inactive for a particular $k$ that
    $y_k<\frac{|\bof_k^H\bu_k|^2}{\tau\sigma^2+\tau\sum_{i=1, i\neq k}^K\frac{|\bof_k^H\bu_i|^2}{y_i}}$,
then another feasible point can be obtained by replacing $y_k$ with $y_k'$ where
$y_k':=\frac{|\bof_k^H\bu_k|^2}{\tau\sigma^2+\tau\sum_{i=1, i\neq k}^K\frac{|\bof_k^H\bu_i|^2}{y_i}}>y_k$.
This is because all the constraints of (\ref{C4}) are satisfied:
$\sum_{j=1}^{Z} |a_{kj}|^2 x_j-\eta y_k'<\sum_{j=1}^{Z}|a_{kj}|^2 x_j-\eta y_k\le 0$,
and for any $j\neq k$,
$    y_j\le\frac{|\bof_j^H\bu_j|^2}{\tau\sigma^2+\tau\left(\sum_{i=1, i\neq j, i\neq k}^K\frac{|\bof_j^H\bu_i|^2}{y_i}+\frac{|\bof_j^H\bu_k|^2}{y_k}\right)}
    <\frac{|\bof_j^H\bu_j|^2}{\tau\sigma^2+\tau\left(\sum_{i=1, i\neq j, i\neq k}^K\frac{|\bof_j^H\bu_i|^2}{y_i}+\frac{|\bof_j^H\bu_k|^2}{y_k'}\right)}$.
%\end{split}
%\end{equation}
Next, we note that the new feasible point $\{x_1,\cdots,x_Z,y_1,\cdots,y_{k-1},y_k',y_{k+1},\cdots,y_K\}$ achieves the lower value of objective function than $\{x_1,\cdots,x_Z, y_1,\cdots,y_K\}$, since
$\sum_{j=1}^{Z}\phi_j(\frac{1}{x_j}-\sigma^2)+\sum_{k=1}^K \frac{1}{y_k}$ is strictly decreasing with $y_k$.
Therefore, the optimal $\{y_k: k=1,\cdots,K\}$ must be achieved when the last constraints for all $k=1,\cdots,K$ are active, which means there are $K$ variables and $K$ equations for the optimal
$    y_k=\frac{|\bof_k^H\bu_k|^2}{\tau\sigma^2+\tau\sum_{i=1, i\neq k}^K\frac{|\bof_k^H\bu_i|^2}{y_i}},\quad k=1,\cdots,K$,
%\end{equation}
which is equivalent to $\bDelta^H\bp+\sigma^2\bone_{K\times 1}=\bzero$. Thus, we can solve the optimal $\bp$ directly in closed form as $\bp=-\sigma^2(\bDelta^H)^{-1}\bone_{K\times 1}\succeq\bzero$. Substituting the optimal $\bp$ to (\ref{C1}) and using $x_j$ as variables instead of $\lambda_j$, the obtained problem is convex and then can be solved.

\section{Proof of Lemma \ref{lemma_equivalent}}\label{appen_equivalence}
First, it can be readily shown that (\ref{solvable_problem}) and (\ref{equivalent_problem}) are equivalent if $C_k=C$ for all $k$. Then it is suffices to show that $C_k=C$ holds for the proposed scheme when $L\ge K+Z$.
Note that in Lemma \ref{lemma2} of the paper, we have shown that $\bB^H\bSigma_{\textrm{opt}}=\bzero$, where $\bSigma_{\textrm{opt}}$ is the optimal $\bSigma$ in (\ref{solvable_problem}). Therefore, $\SINR_k(\bp,\bSigma_{\textrm{opt}})$ is only a function of $\bp$. Furthermore, from Theorem \ref{theorem1} we know that the optimal $\bp$ in (\ref{solvable_problem}) can be computed by
$    \bp_{\textrm{opt}}=-\sigma^2\left(\bDelta^H\right)^{-1}\bone_{K\times 1}$.
Substituting $\bp_{\textrm{opt}}$ to $\SINR_k(\bp,\bSigma_{\textrm{opt}})$, one can readily verify that $\SINR_k(\bp_{\textrm{opt}},\bSigma_{\textrm{opt}})=\tau$, which means $C_k=C$.

\section{Proof of Lemma \ref{lemma6}}\label{proof_lemma6}
When $\Ptot\rightarrow\infty$, the third constraint of (\ref{alternating_problem}) is relaxed. Therefore, we can rewrite (\ref{alternating_problem}) as:
\begin{equation}\label{temp_lemma6}
  \begin{split}
    \min_{\bGamma,\{\bc_k\},\{x_j\ge 0\},}&\quad\max_{k}\left\{p_k\sum_{j=1}^{Z}\frac{|a_{kj}|^2}{\sigma^2+x_j^2}\right\}\quad\\
      \st\quad&\bG^H\bGamma^H=[\diag\{x_1,x_2,\cdots,x_{Z}\},\bzero]^T,\\
      &\bb_k^H\bGamma^H=\bc_k^H,\quad k=1,\cdots,K,
      %&\bDelta^H\bp+[\|\bc_1\|^2,\cdots,\|\bc_K\|^2]^T+\sigma^2\bone\preceq\bzero,\quad\bp\succeq\bzero.
  \end{split}
\end{equation}
where $p_k=\bdelta_k^H[\|\bc_1\|+\sigma^2,\cdots,\|\bc_K\|+\sigma^2]^T$. Note that $\max_{k}\left\{p_k\sum_{j=1}^{Z}\frac{|a_{kj}|^2}{\sigma^2+x_j^2}\right\}$ is a decreasing function of $\{x_j\}$. Thus, by increasing $x_j$, the objective function of (\ref{temp_lemma6}) can always be decreased, which means if $x_j\rightarrow\infty$ for all $j=1,\cdots, Z$ is feasible, then $x_j\rightarrow\infty$ for all $j=1,\cdots, Z$ must be optimal. In order to prove that $x_j\rightarrow\infty$ for all $j=1,\cdots, Z$ is feasible, without loss of generality, we can instead prove that for any given feasible point $\bGamma$ and $\{x_1,x_2,\cdots,x_Z\}$, one can always find another feasible point $\bGamma'$ and $\{x_1, x_2,\cdots,x_{j-1},x_j',x_{j+1},\cdots,x_Z\}$ which can achieve a lower or equal objective value compared to than $\{x_1,\cdots,x_Z\}$. Note that since the objective function of (\ref{temp_lemma6}) is non-decreasing in $x_j$, we only need to prove that the solution of $\bGamma'$ exists for $\{x_1, x_2,\cdots,x_{j-1},x_j',x_{j+1},\cdots,x_Z\}$, where $x_j'>x_j$. This is equivalent to proving that there exists $\bGamma'$ which satisfies $\bG^H(\bGamma'-\bGamma)^H=[\diag\{0,\cdots,0,x_j'-x_j,0,\cdots,0\},\bzero]^T$. Since $\bGamma'\in\mathbb{C}^{Z\times L}$, $\bG\in\mathbb{C}^{L\times Z}$, and $L\ge Z$, the solution of $\bG^H(\bGamma'-\bGamma)^H=[\diag\{0,\cdots,0,x_j'-x_j,0,\cdots,0\},\bzero]^T$ must exist.

\section{Proof of Lemma \ref{lemma3}}\label{proof_lemma3}
Denote the optimal $x_j: j=1,\cdots,Z$ of (\ref{alternating_problem}) as $x_j^*$ and the optimal $\eta$ as $\eta_{\textrm{opt}}$. Then it is easy to prove that at least one of the following $K$ constraints must be active: $p_k\sum_{j=1}^{Z}\frac{|a_{kj}|^2}{\sigma^2+(x_j^*)^2}\le\eta_{\textrm{opt}}, \quad k=1,\cdots,K$.
Thus, we can write $\eta_{\textrm{opt}}$ as
    $\eta_{\textrm{opt}}=\max_{k} \left\{p_k\sum_{j=1}^{Z}\frac{|a_{kj}|^2}{\sigma^2+(x_j^*)^2}\right\}$.
Since in the first iteration of Step 1 in (\ref{alternating_step1}) we set $\btc=\bzero$ as the initial setting, which is optimal when $L\ge K+Z$, denoting the resulting $\eta$ of (\ref{alternating_step1}) with $\btc=\bzero$ as $\eta_{\textrm{init}}$, one can easily prove that
        $\eta_{\textrm{sub}}\le\eta_{\textrm{init}}\le\max_{k} \left\{p_k\sum_{j=1}^{Z}\frac{|a_{kj}|^2}{(x_j^*)^2}\right\}$,
where $\eta_{\textrm{sub}}$ denotes the obtained $\eta$ by the proposed suboptimal algorithm. Then we have
$\eta_{\textrm{sub}}-\eta_{\textrm{opt}}\le\max_{k} \left\{p_k\left(\sum_{j=1}^{Z}\frac{|a_{kj}|^2}{(x_j^*)^2}-\sum_{j=1}^{Z}\frac{|a_{kj}|^2}{\sigma^2+(x_j^*)^2}\right)\right\}
    =\max_{k} \left\{p_k\left(\sum_{j=1}^{Z}\frac{\sigma^2|a_{kj}|^2}{(x_j^*)^2[\sigma^2+(x_j^*)^2]}\right)\right\}$.
When $\Ptot\rightarrow\infty$, we have $\eta_{\textrm{sub}}-\eta_{\textrm{opt}}\rightarrow 0$ since $x_j^*\rightarrow\infty$. Thus, the proposed suboptimal algorithm is asymptotically optimal when $L\le K+Z$ and $\Ptot\rightarrow\infty$.

\section{Proof of Lemma \ref{lemma5}}\label{proof_lemma5}
To prove Lemma \ref{lemma5}, we use the same methodology as in Appendix \ref{proof_lemma3}. Denote the optimal $\eta$ of (\ref{alternating_step1}) with $\btc=\bzero$ as $\eta_{\textrm{init}}$, the optimal $\eta$ of the proposed suboptimal algorithm as $\eta_{\textrm{sub}}$, and the asymptotic optimal $\eta$ of (\ref{alternating_problem}) when $\Ptot\rightarrow\infty$ as $\eta_{\textrm{asy}}$. Then it is easy to prove that $\eta_{\textrm{asy}}\le\eta_{\textrm{sub}}\le\eta_{\textrm{init}}$.
When $\bB\rightarrow\bzero$, the problem (\ref{alternating_problem}) turns to
\begin{equation}\label{E1}
  \begin{split}
    &\min_{\bGamma,\{\bc_k\},\{x_j\ge 0\},\eta}\eta\quad\\
      &\st\quad\bG^H\bGamma^H=[\bLambda^{1/2},\bzero]^T\bV^H, \\ &p_k\sum_{j=1}^{Z}\frac{|a_{kj}|^2}{\sigma^2+x_j^2}\le\eta,k=1,\cdots,K,\\
      &\tr\{\bGamma^H\bGamma\}-\|(\bDelta^H)^{-1}[\|\bc_1\|+\sigma^2,\|\bc_2\|+\sigma^2,\cdots,\|\bc_K\|+\sigma^2]^T\|_1\\
      &\quad\le\Ptot,
     % &\bb_k^H\bGamma^H=\bc_k^H,\quad k=1,\cdots,K.
      %&\bDelta^H\bp+[\|\bc_1\|^2,\cdots,\|\bc_K\|^2]^T+\sigma^2\bone\preceq\bzero,\quad\bp\succeq\bzero.
  \end{split}
\end{equation}
from which one can easily prove that $\bc_k=\bzero$ is optimal. One the other hand,
when $\bB\rightarrow\bzero$, the problem (\ref{alternating_step1}) with $\btc=\bzero$ turns to
    \begin{equation}\label{}
    \begin{split}
    \min_{\{x_j\},\bGamma,\eta}&\eta\quad\\
      \st\quad &\bG^H\bGamma^H=[\diag\{x_1,\cdots,x_{Z}\},\bzero]^T,\quad \\
      &x_j\ge 0, j=1,\cdots,Z,\\
      &\tr\{\bGamma^H\bGamma\}\le\Ptot-\sum_{k=1}^K\bdelta_k^H(\sigma^2\bone),\\
      &\sum_{j=1}^{Z}\frac{|a_{kj}|^2}{x_j^2}\le\frac{\eta}{\bdelta_k^H(\sigma^2\bone)},\quad k=1,\cdots,K,
    %  &\left(\bb_1^H\bGamma^H\bGamma\bb_1,\cdots,\bb_K^H\bGamma^H\bGamma\bb_K\right)^T\preceq\btc,\\
    \end{split}
    \end{equation}
which is the same problem as (\ref{E1}).
Thus, when $\Ptot\rightarrow\infty$, the problem (\ref{alternating_problem}) converges to the problem of (\ref{alternating_step1}), which means $\eta_{\textrm{asy}}\rightarrow\eta_{\textrm{init}}$. Since $\eta_{\textrm{asy}}\le\eta_{\textrm{sub}}\le\eta_{\textrm{init}}$, we can conclude that $\eta_{\textrm{sub}}\rightarrow\eta_{\textrm{asy}}$ as $\bB\rightarrow\bzero$. Thus, the proposed suboptimal algorithm is asymptotically optimal when $\bB\rightarrow\bzero$.

\bibliographystyle{IEEEtran}
\bibliography{CJ_multiuser}

\begin{IEEEbiography}[{\includegraphics[width=1in,height=1in,clip,keepaspectratio]{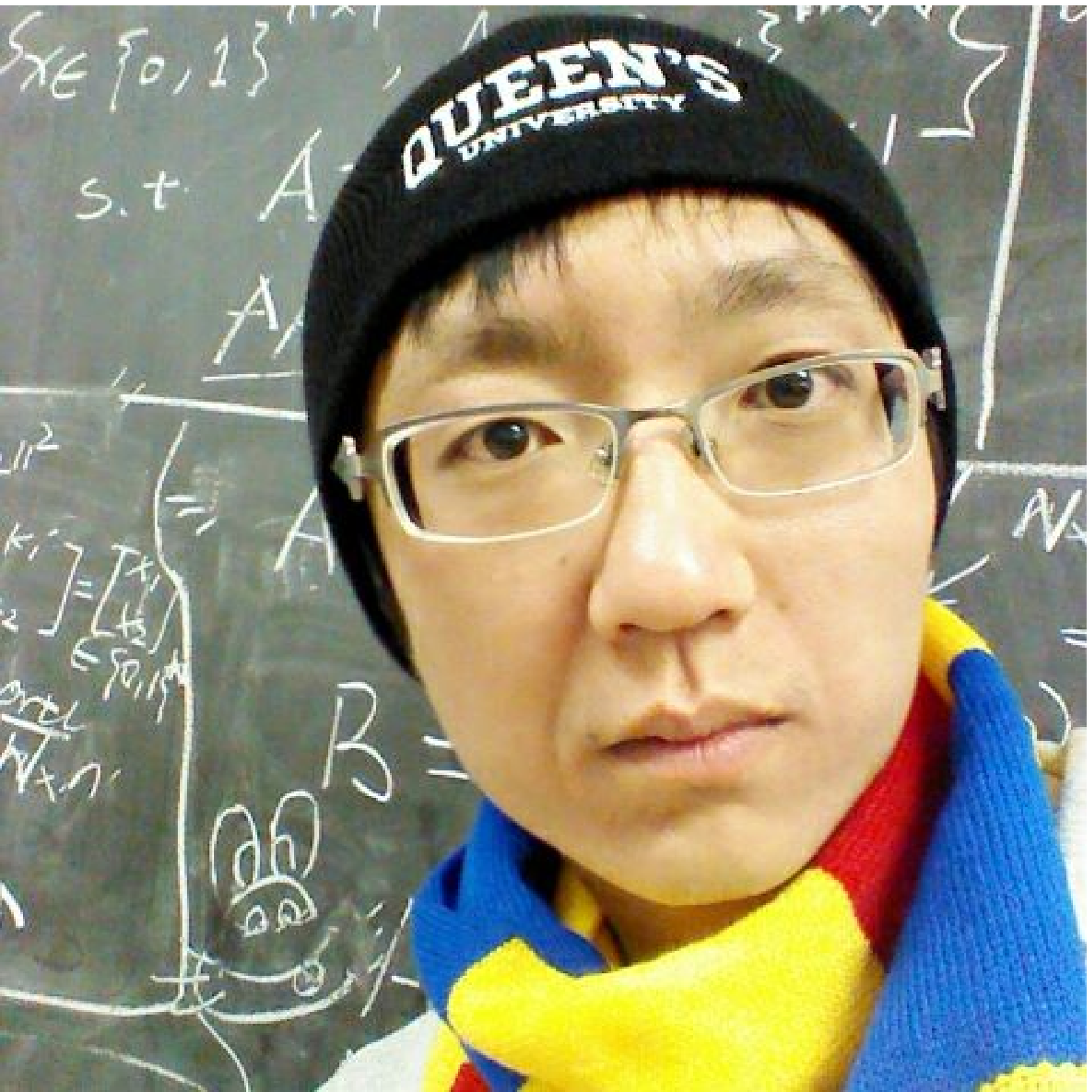}}]{Jun Yang}
received the B. Eng. degree in electrical engineering from Tsinghua University, China, the D. Eng. degree in electrical engineering from Chinese Academy of Sciences, the M. Sc. degree in applied mathematics from Queen's University, Canada. From Dec. 2004 to Sept. 2005, he worked at Bell Labs Research China, Lucent Technologies as a research intern. From Nov. 2010 to July 2012 and from Dec. 2012 to Aug. 2013, he was a postdoctoral fellow at Department of Electrical and Computer Engineering, Queen's University. From Sept. 2013 to Aug. 2014, he was with Department of Mathematics and Statistics, Queen's University. Since Sept. 2014, he has been with Department of Statistical Sciences, University of Toronto. His current research interests include applied probability, statistical learning and signal processing.
\end{IEEEbiography}

\begin{IEEEbiography}[{\includegraphics[width=1in,height=1.25in,clip,keepaspectratio]{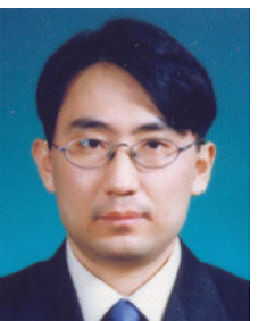}}]{Il-Min Kim}
(SM'06) received the B.Sc. degree in electronics engineering from Yonsei University, Seoul, Korea, in 1996, and the M.S. and Ph.D. degrees in electrical engineering from the Korea Advanced Institute of Science and Technology (KAIST), Taejon, in 1998 and 2001, respectively. From October 2001 to August 2002, he was with the Department of Electrical Engineering and Computer Sciences, MIT, Cambridge, and from September 2002 to June 2003, he was with the Department of Electrical Engineering, Harvard University, Cambridge, MA, as a Postdoctoral Research Fellow. In July 2003, he joined the Department of Electrical and Computer Engineering, Queens University, Kingston, Canada, where he is currently an Associate Professor. His research interests include cognitive radio, wireless bidirectional communications, cooperative diversity networks, physical layer security, CoMP, cross-layer optimization, and network coding. Dr. Kim was as an Editor for the IEEE TRANSACTIONS ON WIRELESS COMMUNICATIONS from 2005 to 2011. He is currently serving as an Editor for the IEEE WIRELESS COMMUNICATIONS LETTERS and as an Editor for the Journal of Communications and Networks (JCN).
\end{IEEEbiography}

\begin{IEEEbiography}[{\includegraphics[width=1in,height=1.25in,clip,keepaspectratio]{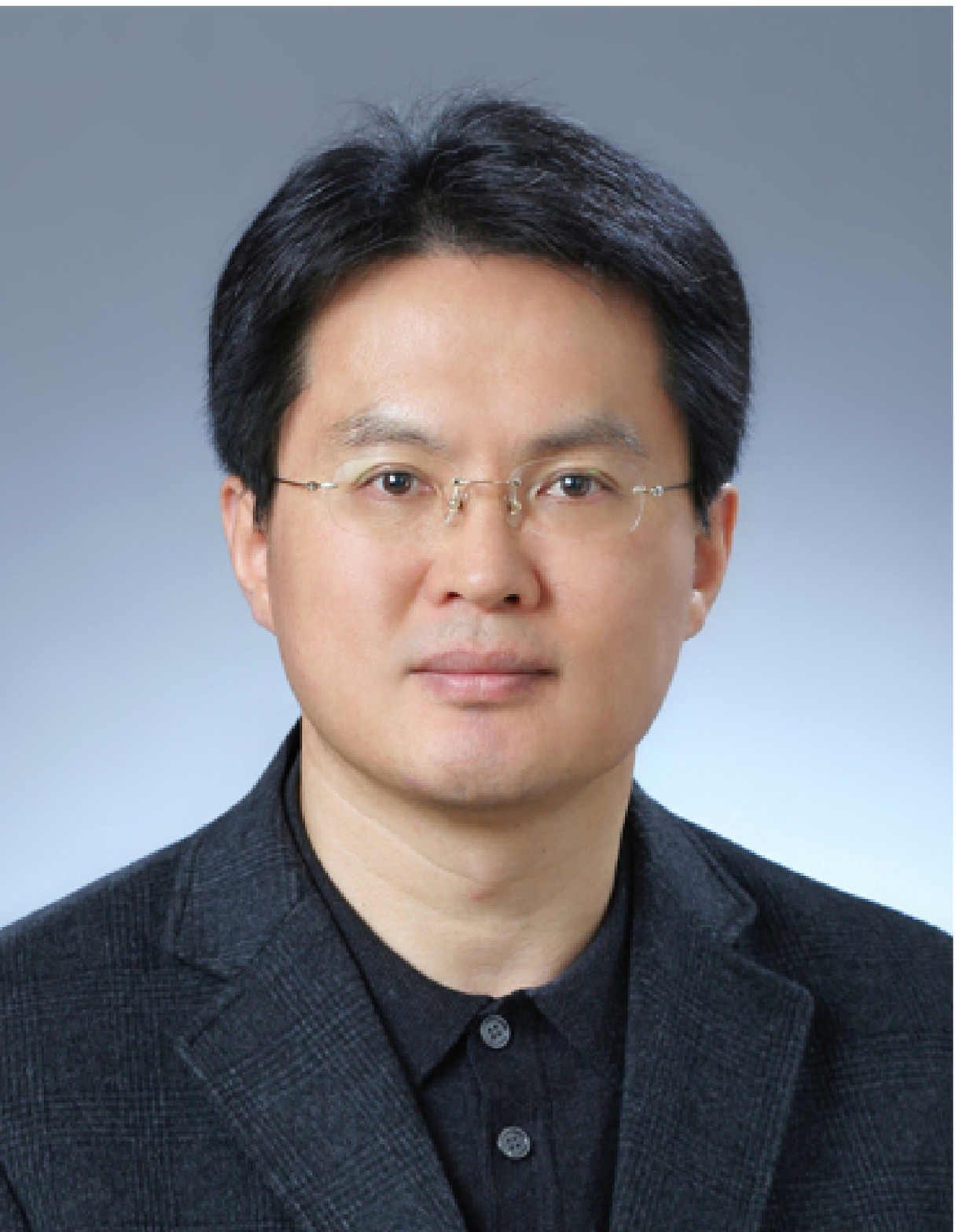}}]{Dong In Kim}
(S'89--M'91--SM'02) received the Ph.D. degree in electrical engineering from the University of Southern California, Los Angeles, in 1990. He was a tenured Professor with the School of Engineering Science, Simon Fraser University, Burnaby, British Columbia, Canada. Since 2007, he has been with Sungkyunkwan University (SKKU), Suwon, Korea, where he is currently a Professor with the College of Information and Communication Engineering. Recently he was awarded the Engineering Research Center (ERC) for Energy Harvesting Communications. Dr. Kim has served as an Editor and a Founding Area Editor of Cross-Layer Design and Optimization for the IEEE Transactions on Wireless Communications from 2002 to 2011. From 2008 to 2011, he served as the Co-Editor-in-Chief for the Journal of Communications and Networks. He is currently the Founding Editor-in-Chief for the IEEE Wireless Communications Letters and has been serving as an Editor of Spread Spectrum Transmission and Access for the IEEE Transactions on Communications since 2001.
\end{IEEEbiography}

\end{document}